\RequirePackage{fix-cm}
\documentclass[smallextended]{svjour3}       
\smartqed  
\usepackage{graphicx}
\usepackage{mathptmx}      
\usepackage{amsfonts}
\usepackage{amssymb}
\usepackage{amsmath}

\journalname{Int J Game Theory}
\begin{document}

\title{Characterizing Permissibility, Proper Rationalizability, and Iterated
Admissibility by Incomplete Information\thanks{The author would like to express her gratitude to Andr\'{e}s Perea for his
thorough reading of the early versions of this paper and his numerous
valuable comments and suggestions. She also would like to thank J\'{a}nos
Flesch, Funaki Yukihiko, Christian Bach, Abraham Neyman, Zsombor Z. M\'{e}%
der, and Dmitriy Kvasov for their valuable discussions and encouragements.
She thanks all teachers and students in the 4th Epicenter Spring Course on
Epistemic Game Theory in Maastricht University for their inspiring teaching
and stimulating discussions. She gratefully acknowledge the support of
Grant-in-Aids for Young Scientists (B) of JSPS No.17K13707 and Grant for
Special Research Project No. 2017K-016 of Waseda University.}
}

\titlerunning{Characterization by Incomplete Information}        

\author{Shuige Liu}


\institute{Shuige Liu \at
              Faculty of Political Science and Economics, Waseda University, 1-6-1
Nishi-Waseda, Shinjuku-Ku, 169-8050, Tokyo, Japan \\
              \email{shuige\_liu@aoni.waseda.jp} 
}

\date{Received: date / Accepted: date}

\maketitle

\begin{abstract}
We characterize three interrelated concepts in epistemic game theory:
permissibility, proper rationalizability, and iterated admissibility. We
define the lexicographic epistemic model for a game with incomplete
information. Based on it, we give two groups of characterizations. The first
group characterizes permissibility and proper rationalizability. The second
group characterizes permissibility in an alternative way and iterated
admissibility. In each group, the conditions for the latter are stronger
than those for the former, which corresponds to the fact that proper
rationalizability and iterated admissibility are two (compatible)
refinements of permissibility within the complete information framework. The
intrinsic difference between the two groups are the role of rationality: the
first group does not need it, while the second group does
\end{abstract}

\section{Introduction}
\label{sec:int}

The purpose of noncooperative game theory is to study an individual's
decision making in an interactive situation. Since there one's payoff is not
completely determined by her own choice, decision making requires her to
form a belief on every other participant's choice, on every other
participant's belief on every other's choice, and so on. Studying the
structure of those belief hierarchies and choices supported by a belief
hierarchy satisfying some particular conditions opened up a field called 
\emph{epistemic game theory.} See Perea (2002) for a textbook on this
field.

In epistemic game theory, various concepts have been developed to describe
specific belief structures. One is \emph{lexicographic belief} (Blume et al. 
1991a, 1991b). A lexicographic belief describes a player's
subjective conjecture about the opponents' behavior by a sequence of
probability distributions over other participants' choices and types, which
is different from the adoption of a single probability distribution in a
standard probabilistic belief. The interpretation of a lexicographic belief
is that every choice-type pair in the sequence is considered to be possible,
while a pair occurring ahead in the sequence is deemed \emph{infinitely more
likely} than one occurring later. Several concepts have been developed by
putting various conditions on lexicographic beliefs intended to capture
different styles of reasoning about the opponents' behavior. Permissibility,
proper rationalizability, and iterated admissibility are three important and
interrelated concepts among them.

\emph{Permissibility} originated from Selten (1975)'s perfect
equilibrium. It is defined and studied from the epistemic viewpoint by using
lexicographic belief hierarchy in Brandenburger (1992) (an alternative
approach without using lexicographic belief is given by B\"{o}rgers, 1994). Permissibility is based on two notions: \emph{caution} and \emph{%
primary belief in the opponents' rationality}. A lexicographic belief is
said to be cautious if it does not exclude any choice of the opponents; it
is said to primarily believe in the opponents' rationality if its first
level belief only deems possible those choice-type pairs where the choice is
optimal under the belief of the paired type (Perea, 2012).

\emph{Proper rationalizability} originated from Myerson (1978)'s proper
equilibrium which is intended to be a refinement of perfect equilibrium. It
is defined and studied in Schuhmacher (1999) and Asheim (2001) as
an epistemic concept. Proper rationalizability shares with permissibility
the notion of caution, while, instead of primary belief in the opponents'
rationality, it is based on a stronger notion called \emph{respecting the
opponents' preferences} which means that a \textquotedblleft
better\textquotedblright\ choice always occurs in front of a
\textquotedblleft worse\textquotedblright\ choice in a lexicographic belief.

\emph{Iterated admissibility} originated from the study of iteratively
undominated choices (Bernheim, 1984, Pearce, 1984, Samuelson, 1992). In Brandenburger et al. (2008) and Perea (2012) it is
explored as an epistemic concept by using lexicographic belief hierarchy. It
is based on caution and a notion called \emph{assumption of rationality}, which is also stronger than primary belief in the opponents' rationality. A
lexicographic belief is said to assume the opponents' rationality if every
\textquotedblleft good\textquotedblright\ choice occurs and is always
located in front of any \textquotedblleft bad\textquotedblright\ one. Here
good means that a choice of the opponent can be supported by a cautious belief of her, and bad means the opposite.

We illustrate the three concepts by two examples.\smallskip \newline
\textbf{Example \ref{sec:int}.1 (Permissibility and proper rationalizability)%
} Consider a game where player $1$ has strategies $A$ and $B$ and player $2$
has strategies $C,D,$ and $E.$ Player $2$'s utility function $u_{2}$ is illustrated in Table \ref{tab:1}.%
\begin{table}
\label{tab:1}
\caption{A two-person game for permissibility and proper rationalizability}
\begin{tabular}{llll}
\hline \noalign{\smallskip}
$u_{2}$ & $C$ & $D$ & $E$ \\ 
\noalign{\smallskip} \hline \noalign{\smallskip}
$A$ & $3$ & $2$ & $1$ \\ 
$B$ & $3$ & $2$ & $1$ \\ 
\noalign{\smallskip} \hline
\end{tabular}%
\end{table}
Consider a lexicographic belief of player $1$ on player $2$'s choices. Caution requires that all three choices of player 2 occur in that belief.
Since $C$ is player $2$'s most preferred choice, primary belief in player $2$%
's rationality requires that only choice $C$ can be put in the first level.
Further, since player 2 prefers $C$ to $D$ and $D$ to $E$, a lexicographic
belief of player 1 respecting 2's preferences should deem $C$ infinitely
more likely than $D$ and $D$ infinitely more likely than $E,$ that is, put $%
C $ before $D$ and $D$ before $E$.\smallskip \newline
\textbf{Example \ref{sec:int}.2 (Iterated admissibility) }Consider $u_{2}$ illustrated in Table 2 which is obtained from the one in Example \ref{sec:int}.1
by changing $u_{2}(B,E)$ from $1$ to $4$. %
\begin{table}
\label{tab:2}
\caption{A two-person game for iterated admissibility}
\begin{tabular}{llll}
\hline \noalign{\smallskip}
$u_{2}$ & $C$ & $D$ & $E$ \\ 
\noalign{\smallskip} \hline \noalign{\smallskip}
$A$ & $3$ & $2$ & $1$ \\ 
$B$ & $3$ & $2$ & \underline{$4$}\\ 
\noalign{\smallskip} \hline
\end{tabular}%
\end{table}
Here, $C$ could be optimal for player 2 if she believes that player 1 will
use choice $A,$ and $E$ could be optimal if she believes player 1 would use
choice $B,$ while $D$ could never be optimal whatever player 2 believes
about player 1's choices. Hence, a lexicographic belief of player 1 assuming
player 2's rationality should deem $C$ and $E$ infinitely more likely than $%
D.$\smallskip

One motivation for developing lexicographic belief is to alleviate the
tension between caution and rationality (Blume et al., 1991a,
Brandenburger, 1992, B\"{o}rgers, 1994, Samuelson, 1992, B%
\"{o}rgers and Samuelson, 1994). All the three concepts tried to solve
the tension by sacrificing rationality in different ways. Indeed,
permissibility requires that the first level belief contains only rational
choices, proper rationalizability requires that choices should be ordered
according to the level of rationality, and iterated admissibility requires
that choices which could be rational to be put in front of those which can
never be. However, due to caution all allow occurrences of irrational
choices. This sacrifice of rationality brought conceptual inconvenience
since rationality is a basic assumption in game theory and is reasonable to
be adopted as a criterion for each player's belief.

There is an approach which solves the tension without sacrificing
rationality: using an incomplete information framework. That is, instead of
considering the uncertainty about the opponents' rationality within a
complete information framework, we take the uncertainty about the opponents'
utility functions and consider types within the incomplete information
framework. Then the occurrence of a irrational choice can be explained as
that the \textquotedblleft real\textquotedblright\ utility function of an
opponent is different from the original one. Permissibility, proper
rationalizability, and iterated admissibility can all be characterized
within an incomplete information framework. This is the basic idea of this
paper.

We use the above examples to explain this idea.\smallskip \newline
\textbf{Example \ref{sec:int}.1 (Continued) }As mentioned before, though
only choice $C$ is rational for player 2, caution requires $D$ and $E$ to
occur in player 1's belief as well. In a complete information framework, the
occurrences of $D$ and $E$ are explained by player 2's irrationality (i.e.,
\textquotedblleft trembling hand\textquotedblright ). In contrast, within an
incomplete information framework they are explained by the possibility that
the \textquotedblleft real\textquotedblright\ utility function of player 2
is not $u_{2}$ but $v_{2}$ or $v_{2}^{\prime }$ in Table 3.
\begin{table}
\label{tab:3}
\caption{Alternative utility functions for player $2$}
\begin{tabular}{llll}
\hline \noalign{\smallskip}
$v_{2}$ & $C$ & $D$ & $E$ \\ 
\noalign{\smallskip} \hline \noalign{\smallskip}
$A$ & $2$ & $3$ & $1$ \\ 
$B$ & $2$ & $3$ & $1$\\ 
\noalign{\smallskip} \hline
\end{tabular}\quad \quad \quad
\begin{tabular}{llll}
\hline \noalign{\smallskip}
$v_{2}^{\prime}$ & $C$ & $D$ & $E$ \\ 
\noalign{\smallskip} \hline \noalign{\smallskip}
$A$ & $2$ & $1$ & $3$ \\ 
$B$ & $2$ & $1$ & $3$\\ 
\noalign{\smallskip} \hline
\end{tabular}%
\end{table}
Choice $D$ is optimal in $v_{2}$ and $E$ is optimal in $v_{2}^{\prime }.$ In
this manner, uncertainty about the opponent's rationality within a complete
information framework is transformed into uncertainty about the opponent's
real utility function within an incomplete information framework. It can be
seen that primary belief in the opponent's rationality in complete
information framework is equivalent to the condition that one deems $u_{2}$
or a utility function \textquotedblleft very similar\textquotedblright\ to $%
u_{2}$ infinitely more likely to be the real utility function of player 2
than $v_{2}$ and $v_{2}^{\prime }$, and respecting the opponent's
preferences is equivalent to the condition that those alternative utility
functions should be ordered by their \textquotedblleft
similarity\textquotedblright\ to $u_{2}$.\smallskip \newline
\textbf{Example \ref{sec:int}.2 (Continued) }Since both $C$ and $E$ can be
optimal in the original game, we only need a game to support choice $D.$
Consider that the \textquotedblleft real\textquotedblright\ utility function
for player $2$ is $w_{2}$ illustrated in Table 4.%
\begin{table}
\label{tab:4}
\caption{An alternative utility function for player $2$}
\begin{tabular}{llll}
\hline \noalign{\smallskip}
$w_{2}$ & $C$ & $D$ & $E$ \\ 
\noalign{\smallskip} \hline \noalign{\smallskip}
$A$ & $3$ & $5$ & $4$ \\ 
$B$ & $3$ & $5$ & $1$\\ 
\noalign{\smallskip} \hline
\end{tabular}%
\end{table}
Choice $D$ is optimal in $w_{2}.$ It can be seen that assumption of player
2's rationality corresponds to that those utility function different from
the original one to be less likely to the original one.\smallskip

In this paper, we study the equivalences between conditions in complete and
incomplete information models for 2-person static games and provide
characterizations of permissibility, proper rationalizability, and
assumption of rationality. First, we define the lexicographic epistemic
model of a game with incomplete information. Then we give our
characterization results which are separated into two groups. In the first
group, we characterize permissibility and proper rationalizability, showing
that a choice is permissible (properly rationalizable) within a complete
information framework if and only if it is optimal for a belief hierarchy
within the corresponding incomplete information framework that expresses
common full belief in caution, primary belief in the opponent's utilities
nearest to the original utilities (the opponent's utilities are centered
around the original utilities), and a best (better) choice is supported by
utilities nearest (nearer) to the original ones.

In the second group, we characterize permissibility in an alternative way
and iterated admissibility, showing that a choice is permissible
(iteratively admissible) within a complete information model if and only
there is a belief hierarchy within the corresponding incomplete information
framework that expresses common full belief of caution, rationality, and
primary belief in the original utilities (prior assumption of the original
utilities and every good choice is supported). We will further show in
Section \ref{sec:cr}.5 that caution can be weakened in this group of
characterizations.

Both proper rationalizability and iterated admissibility are refinements of
permissibility (Perea, 2012). This can also be seen in the
characterizations within the incomplete information models: in each group,
each conditions for the latter is stronger than its counterpart for the
former. The intrinsic difference between the two groups are the role of
rationality: the first group does not need it , while the second group does.
Nevertheless, we can construct a belief hierarchy which satisfies
rationality and conditions in the first group. This corresponds to the fact
that, within an complete information framework, it is always possible to
have a belief hierarchy satisfying the conditions of proper
rationalizability as well as those of iterated admissibility.\smallskip

This paper is not the first one characterizing concepts in epistemic game
theory within the incomplete information framework. Perea and Roy (2017)
characterized $\varepsilon $-proper rationalizability in this approach by
using a standard epistemic model without lexicographic beliefs. They showed
that a type in a standard epistemic model with complete information
expresses common full belief in caution and $\varepsilon $-trembling
condition if and only if there is a type in the corresponding model with
incomplete information sharing the same belief hierarchy with it which
expresses common belief in caution, $\varepsilon $-centered belief around
the original utilities $u$, and belief in rationality under the closest
utility function. Since each properly rationalizable choice is the limit of
a sequence of $\varepsilon $-proper rationalizable ones, the conditions
adopted in their characterizations are very useful for us. Two conditions in
our characterization of proper rationalizability, that is, caution and $u$%
-centered belief, are faithful translations of their conditions into
lexicographic model. However, the most critical condition in their
characterization, that is, belief in rationality under the closest utility
function, is impossible to be adopted here. The reason is, as will be shown
in Section \ref{sec:asu}.1, that a nearest utility function making a choice
optimal does not always exist in lexicographic models. This is a salient
difference between standard probabilistic beliefs and lexicographic ones. We
define a weaker condition called \textquotedblleft a better choice is
supported by utilities nearer to the original one\textquotedblright\ and
show that it can be used to characterize proper rationalizability.

Another essential difference between Perea and Roy (2017) and this
paper is in the way of proof. Equivalence of belief hierarchies generated by
types in models with complete and incomplete informations and type morphisms
(B\"{o}ge and Eisele, 1979, Heifetz and Samet, 1998, Perea and
Kets, 2016) play important roles in Perea and Roy (2017)'s proof.
In contrast, our proofs are based on constructing a specific correspondence
between the two models. We show that conditions in a type of one model
implies that appropriate conditions are satisfied in the corresponding type
in the constructed model. Equivalence of hierarchies follows directly by
construction. Our construction can also be used to prove Perea and Roy (2017)'s Theorem 6.1. Further, as will be discussed in Section \ref{sec:cr}%
.3, our construction shows that rationality is separable from other
conditions in characterizing proper rationalizability. This confirms that
the consistency of caution and rationality within an incomplete information
framework.

Our results, as well as Perea and Roy (2017)'s, also provide insights
in decision theory and general epistemology. They imply that any choice
permissible, properly rationalizable, or iteratively admissible within a
complete information framework is also optimal for a belief satisfying
corresponding conditions within an incomplete information framework, and
vice versa. In other words, by just looking at the outcome, it is impossible
to know the accurate epistemic situation behind the choice, that is, whether
it is because of players' uncertainty about the opponents' rationality or
uncertainty about what are the real utilities of the opponents.

The rest of the paper is organized as follows. Section \ref{sec:mac} defines
permissibility, proper rationalizability, and iterated admissibility in
epistemic models with complete information. Section \ref{sec:asu} introduces
the lexicographic epistemic model with incomplete information and defines
two groups of conditions on the types in such a model. Section \ref{sec:cce}
gives two groups of characterizations and their proofs. Section \ref{sec:cr}
gives some concluding remarks. Section \ref{sec:pro} contains the proofs of
all lemmas.

\section{Epistemic Model with Complete Information}
\label{sec:mac}

In this section, we give a survey of lexicographic epistemic model with
complete information. Definitions here follow Perea (2012), Chapters 5-7.

Consider a finite 2-person static game $\Gamma =(C_{i},u_{i})_{i\in I}$
where $I=\{1,2\}$ is the set of players, $C_{i}$ is the finite set of
choices and $u_{i}:C_{1}\times C_{2}\rightarrow \mathbb{R}$ is the utility
function for player $i\in I.$ In the following sometimes we denote $%
C_{1}\times C_{2}$ by $C$. We assume that each player has a lexicographic
belief on the opponent's choices, a lexicographic belief on the opponent's
lexicographic belief on her, and so on. This belief hierarchy is described
by a lexicographic epistemic model with types.\smallskip \newline
\textbf{Definition \ref{sec:mac}.1 (Epistemic model with complete
information).} Consider a finite 2-person static game $\Gamma
=(C_{i},u_{i})_{i\in I}$. A finite \emph{lexicographic epistemic model} for $%
\Gamma $ is a tuple $M^{co}=(T_{i},b_{i})_{i\in I}$ where\smallskip \newline
(a) $T_{i}$ is a finite set of types, and\smallskip \newline
(b) $b_{i}$ is a mapping that assigns to each $t_{i}\in T_{i}$ a
lexicographic belief over $\Delta (C_{j}\times T_{j}),$ i.e., $%
b_{i}(t_{i})=(b_{i1},b_{i2},...,b_{iK})$ where $b_{ik}\in \Delta
(C_{j}\times T_{j})$ for $k=1,...,K.$\smallskip

Consider $t_{i}\in T_{i}$ with $b_{i}(t_{i})=(b_{i1},b_{i2},...,b_{iK}).$
Each $b_{ik}$ ($k=1,...,K$) is called $t_{i}$'s \emph{level-}$k$\emph{\
belief}. For each $(c_{j},t_{j})\in C_{j}\times T_{j},$ we say $t_{i}$ \emph{%
deems} $(c_{j},t_{j})$ \emph{possible} iff $b_{ik}(c_{j},t_{j})>0$ for some $%
k\in \{1,...,K\}.$ We say $t_{i}$ \emph{deems} $t_{j}\in T_{j}$ \emph{%
possible} iff $t_{i}$ deems $(c_{j},t_{j})$ possible for some $c_{j}\in
C_{j} $. For each $t_{i}\in T_{i},$ we denote by $T_{j}(t_{i})$ the set of
types in $T_{j}$ deemed possible by $t_{i}$.

A type $t_{i}\in T_{i}$ is \emph{cautious} iff for each $c_{j}\in C_{j}$ and
each $t_{j}\in T_{j}(t_{i}),$ $t_{i}$ deems $(c_{j},t_{j})$ possible. That
is, $t_{i}$ takes into account each choice of player $j$ for every belief
hierarchy of $j$ deemed possible by $t_{i}.$\smallskip

For each $c_{i}\in C_{i}$, let $%
u_{i}(c_{i},t_{i})=(u_{i}(c_{i},b_{i1}).,..,u_{i}(c_{i},b_{iK}))$ where for
each $k=1,...,K,$ $u_{i}(c_{i},b_{ik}):=\Sigma _{(c_{j},t_{j})\in
C_{j}\times T_{j}}b_{ik}(c_{j},t_{j})u_{i}(c_{i},c_{j}),$ that is, each $%
u_{i}(c_{i},b_{ik})$ is the expected utility for $c_{i}$ over $b_{ik}$ and $%
u_{i}(c_{i},t_{i})$ is a vector of expected utilities. For each $%
c_{i},c_{i}^{\prime }\in C_{i}$, we say that $t_{i}$ \emph{prefers} $c_{i}$ 
\emph{to} $c_{i}^{\prime }$, denoted by $u_{i}(c_{i},t_{i})>u_{i}(c_{i}^{%
\prime },t_{i}),$ iff there is $k\in \{0,...,K-1\}$ such that the following
two conditions are satisfied:\smallskip \newline
(a) $u_{i}(c_{i},b_{i\ell })=u_{i}(c_{i}^{\prime },b_{i\ell })$ for $\ell
=0,...,k,$ and\smallskip \newline
(b) $u_{i}(c_{i},b_{i,k+1})>u_{i}(c_{i}^{\prime },b_{i,k+1})$.\smallskip 
\newline
We say that $t_{i}$ \emph{is indifferent between }$c_{i}$ \emph{and }$%
c_{i}^{\prime },$ denoted by $u_{i}(c_{i},t_{i})=u_{i}(c_{i}^{\prime
},t_{i}),$ iff $u_{i}(c_{i},b_{ik})=u_{i}(c_{i}^{\prime },b_{ik})$ for each $%
k=1,...,K.$ It can be seen that this preference relation on $C_{i}$ under
each type $t_{i}$ is a linear order. $c_{i}$ is \emph{rational} (or \emph{%
optimal}) for $t_{i}$ iff $t_{i}$ does not prefer any choice to $c_{i}$. A
type $t_{i}\in T_{i}$ \emph{primarily believes in }$\emph{the}$ \emph{%
opponent's rationality} iff $t_{i}$'s level-1 belief only assigns positive
probability to those $(c_{j},t_{j})$ where $c_{j}$ is rational for $t_{j}.$
That is, at least in the primary belief $t_{i}$ is convinced that $j$
behaves rationally given her belief.\smallskip

For $(c_{j},t_{j}),(c_{j}^{\prime },t_{j}^{\prime })\in C_{j}\times T_{j},$
we say that $t_{i}$ \emph{deems} $(c_{j},t_{j})$ \emph{infinitely more
likely than} $(c_{j}^{\prime },t_{j}^{\prime })$ iff there is $k\in
\{0,...,K-1\}$ such that the following two conditions are
satisfied:\smallskip \newline
(a) $b_{i\ell }(c_{j},t_{j})=b_{i\ell }(c_{j}^{\prime },t_{j}^{\prime })=0$
for $\ell =0,...,k,$ and\smallskip \newline
(b) $b_{i,k+1}(c_{j},t_{j})>0$ and $b_{i,k+1}(c_{j}^{\prime },t_{j}^{\prime
})=0$.\smallskip

A cautious type\textbf{\ }$t_{i}\in T_{i}$ \emph{respects the opponent's
preferences} iff for each $t_{j}\in T_{j}(t_{i})$ and $c_{j},c_{j}^{\prime
}\in C_{j}$ where $t_{j}$ prefers $c_{j}$ to $c_{j}^{\prime },$ $t_{i}$
deems $(c_{j},t_{j})$ infinitely more likely than $(c_{j}^{\prime },t_{j}).$
That is, $t_{i}$ arranges $j$'s choices from the most to the least preferred
for each belief hierarchy of $j$ deemed possible by $t_{i}$. It can be seen
that respect of the opponent's preferences implies primary belief in the
opponent's rationality. Indeed, the former requires that each type of the
opponent deemed possible in the primary belief should only pair with choices
most preferred under that type.\smallskip

Let $P$ be an arbitrary property of lexicographic beliefs. We define
that\smallskip \newline
(a) $t_{i}\in T_{i}$ \emph{expresses }$0$\emph{-fold full belief in} $P$ iff 
$t_{i}$ satisfies $P;$\smallskip \newline
(b) For each $n\in \mathbb{N},$ $t_{i}\in T_{i}$ \emph{expresses }$(n+1)$%
\emph{-fold full belief in} $P$ iff $t_{i}$ only deems possible $j$'s types
that express $n$-fold full belief in $P.$\smallskip

A type $t_{i}$ \emph{expresses common full belief in} $P$ iff it expresses $%
n $-fold full belief in $P$ for each $n\in \mathbb{N}.$\smallskip \newline
\textbf{Definition \ref{sec:mac}.2 (Permissibility and proper
rationalizability)}. Consider a finite lexicographic epistemic model $%
M^{co}=(T_{i},b_{i})_{i\in I}$ for a game $\Gamma =(C_{i},u_{i})_{i\in I}$. $%
c_{i}\in C_{i}$ is \emph{permissible} iff it is rational for some $t_{i}\in
T_{i}$ which expresses common full belief in caution and primary belief in
rationality. $c_{i}$ is \emph{properly rationalizable} iff it is rational
for some $t_{i}\in T_{i}$ which expresses common full belief in caution and
respect of preferences.\smallskip

Since respect of the opponent's preferences implies primary belief in the
opponent's rationality, proper rationalizability implies permissibility,
while the reverse does not hold.\smallskip

A cautious type\textbf{\ }$t_{i}\in T_{i}$ \emph{assumes the }$j$\emph{'s
rationality} iff the following two conditions are satisfied:\smallskip 
\newline
(a) for all of player $j$'s choices $c_{j}$ that are optimal for some
cautious belief, $t_{i}$ deems possible some cautious type $t_{j}$ for which 
$c_{j}$ is optimal;\smallskip \newline
(b) $t_{i}$ deems all choice-type pairs $(c_{j},t_{j})$ where $t_{j}$ is
cautious and $c_{j}$ is optimal for $t_{j}$ infinitely more likely than any
choice-type pairs $(c_{j}^{\prime },t_{j}^{\prime })$ that does not have
this property.\smallskip

Informally speaking, assumption of the opponent's rationality is that $t_{i}$
puts all \textquotedblleft good\textquotedblright\ choices in front of those
\textquotedblleft bad\textquotedblright\ choices.

On the other hand, extending assumption of rationality into $n$-fold for any 
$n\in \mathbb{N}$ is more complicated than $n$-fold full belief. Formally,
consider a finite lexicographic epistemic model $M^{co}=(T_{i},b_{i})_{i\in
I}$ for a game $\Gamma =(C_{i},u_{i})_{i\in I}$. A cautious type $t_{i}\in
T_{i}$ \emph{expresses }$1$\emph{-fold assumption of rationality} iff it
assumes $j$'s rationality. For any $n\in \mathbb{N},$ we say that a cautious
type $t_{i}\in T_{i}$ expresses $(n+1)$\emph{-fold assumption of rationality}
iff the following two conditions are satisfied:\smallskip \newline
(a) whenever a choice $c_{j}$ of player $j$ is optimal for some cautious
type (not necessarily in $M^{co}$) that expresses up to $n$-fold assumption
of rationality, $t_{i}$ deems possible some cautious type $t_{j}$ for player 
$j$ which expresses up to $n$-fold assumption of rationality for which $%
c_{j} $ is optimal;\smallskip \newline
(b) $t_{i}$ deems all choice-type pair $(c_{j},t_{j})$ where $t_{j}$ is
cautious and expresses up to $n$-fold assumption of rationality and $c_{j}$
is optimal for $t_{j}$ infinitely more likely than any choice-type pair $%
(c_{j}^{\prime },t_{j}^{\prime })$ that does not satisfy this property.

We say that $t_{i}$ \emph{expresses common assumption of rationality} iff it
expresses $n$-fold assumption of rationality for every $n\in \mathbb{N}.$%
\smallskip \newline
\textbf{Definition \ref{sec:mac}.3 (Iterated admissibility)}. Consider a
finite lexicographic epistemic model $M^{co}=(T_{i},b_{i})_{i\in I}$ for a
game $\Gamma =(C_{i},u_{i})_{i\in I}$. $c_{i}$ is \emph{iteratively
admissible} iff it is rational for some $t_{i}\in T_{i}$ which expresses
common assumption of rationality.

\section{Epistemic Model with Incomplete Information}
\label{sec:asu}

In this section, we define the lexicographic epistemic model with incomplete
information which is the counterpart of the probabilistic epistemic model
with incomplete information introduced by Battigalli (2003) and
extensively developed in Battigalli and Siniscalchi (2003), (2007), and Dekel and Siniscalchi (2015). We also define conditions on types
in such a model.\smallskip \newline
\textbf{Definition \ref{sec:asu}.1 (Lexicographic epistemic model with
incomplete information)}. Consider a finite 2-person static game form $%
G=(C_{i})_{i\in I}.$ For each $i\in I,$ let $V_{i}$ be the set of utility
functions $v_{i}:C_{1}\times C_{2}\rightarrow \mathbb{R}.$ A \emph{finite
lexicographic epistemic model for }$G$\emph{\ with incomplete information}
is a tuple $M^{in}=(\Theta _{i},w_{i},\beta _{i})_{i\in I}$ where\smallskip 
\newline
(a) $\Theta _{i}$ is a finite set of types,\smallskip \newline
(b) $w_{i}$ is a mapping that assigns to each $\theta _{i}\in \Theta _{i}$ a
utility function $w_{i}(\theta _{i})\in V_{i},$ and\smallskip \newline
(c) $\beta _{i}$ is a mapping that assigns to each $\theta _{i}\in \Theta
_{i}$ a lexicographic belief over $\Delta (C_{j}\times \Theta _{j}),$ i.e., $%
\beta _{i}(\theta _{i})=(\beta _{i1},\beta _{i2},...,\beta _{iK})$ where $%
\beta _{ik}\in \Delta (C_{j}\times \Theta _{j})$ for $k=1,...,K.$\smallskip

Concepts such as \textquotedblleft $\theta _{i}$ deems $(c_{j},\theta _{j})$
possible\textquotedblright\ and \textquotedblleft $\theta _{i}$ deems $%
(c_{j},\theta _{j})$ infinitely more likely than $(c_{j}^{\prime },\theta
_{j}^{\prime })$\textquotedblright\ can be defined in a similar way as in
Section \ref{sec:mac}. For each $\theta _{i}\in \Theta _{i},$ we use $\Theta
_{j}(\theta _{i})$ to denote the set of types in $\Theta _{j}$ deemed
possible by $\theta _{i}$. For each $\theta _{i}\in \Theta _{i}$ and
\thinspace $v_{i}\in V_{i},$ $\theta _{i}^{v_{i}}$ is the auxiliary type
satisfying that $\beta _{i}(\theta _{i}^{v_{i}})=\beta _{i}(\theta _{i})$
and $w_{i}(\theta _{i}^{v_{i}})=v_{i}$.

For each $c_{i}\in C_{i},v_{i}\in V_{i},$ and $\theta _{i}\in \Theta _{i}$
with $\beta _{i}(\theta _{i})=(\beta _{i1},\beta _{i2},...,\beta _{iK}),$
let $v_{i}(c_{i},\theta _{i})=(v_{i}(c_{i},\beta
_{i1}),...,v_{i}(c_{i},\beta _{iK}))$ where $v_{i}(c_{i},\beta _{ik}):=\Sigma _{(c_{j},\theta _{j})\in C_{j}\times \Theta
_{j}}\beta _{ik}(c_{j},\theta _{j})v_{i}(c_{i},c_{j})$ for each $k=1,...,K$,  For each $c_{i},c_{i}^{\prime }\in C_{i}$ and $\theta _{i}\in \Theta _{i},$ we say that $\theta _{i}$ \emph{prefers }$c_{i}$ \emph{to} $c_{i}^{\prime }$ iff $%
w_{i}(\theta _{i})(c_{i},\theta _{i})$ $>w_{i}(\theta _{i})(c_{i}^{\prime
},\theta _{i}).$ As in Section \ref{sec:mac}, this is also the lexicographic
comparison between two vectors. $c_{i}$ is \emph{rational} (or \emph{optimal}%
) for $\theta _{i}$ iff $\theta _{i}$ does not prefer any choice to $c_{i}.$

In the following we define two groups of conditions on types in an epistemic
model with incomplete information, which correspond to the two groups of
characterizations in Section \ref{sec:cce}.

\subsection{The first group of conditions}

\textbf{Definition \ref{sec:asu}.2 (Caution)}. A type $\theta _{i}\in \Theta
_{i}$ is \emph{cautious} iff for each $c_{j}\in C_{j}$ and each $\theta
_{j}\in \Theta _{j}(\theta _{i})$, there is some utility function $v_{j}\in
V_{j}$ such that $\theta _{i}$ deems $(c_{j},\theta _{j}^{v_{j}})$
possible.\smallskip

This is a faithful translation of Perea and Roy (2017)'s definition of
caution in a probabilistic model into a lexicographic one. It is the
counterpart of caution defined within the complete information framework in
Section \ref{sec:mac}; the only difference is that in incomplete information
models we allow different utility functions since $c_{j}$ will be required
to be rational for the paired type.\smallskip

For each $u_{i},v_{i}\in V_{i},$ we define the distance $d(u_{i},v_{i})$
between $u_{i},v_{i}$ by $d(u_{i},v_{i})=[\Sigma _{s\in
S}(u_{i}(s)-v_{i}(s))^{2}]^{1/2}.$ This is the Euclidean distance on $%
\mathbb{R}^{C}.$ We choose it is just out of simplicity. Any distance
satisfying the three conditions in Section 3.3 of Perea and Roy (2017)
also works in our characterization.

A problem here is that utility functions are numerical representations of
preferences, yet the Euclidean distance measures cardinal similarity between
utility functions rather than the similarity between preferences they
represent. For example, though multiplying $u_{i}$ with a positive number
leads to the same preferences represented by $u_{i},$ its Euclidean distance
from $u_{i}$ may be large. In Section \ref{sec:cr}.4 we will define an
ordinal distance on $V_{i}$ and show that the characterizations still hold
under that distance.\smallskip \newline
\textbf{Definition \ref{sec:asu}.3 (Primary belief in utilities nearest to }$%
u$\textbf{\ and }$u$\textbf{-centered belief)}. Consider a static game form $%
G=(C_{i})_{i\in I},$ a\emph{\ }lexicographic epistemic model $M^{in}=(\Theta
_{i},w_{i},\beta _{i})_{i\in I}$ for\emph{\ }$G$\emph{\ }with incomplete
information, and a pair $u=(u_{i})_{i\in I}$ of utility functions.\smallskip 
\newline
\textbf{(3.1)} A type $\theta _{i}\in \Theta _{i}$ \emph{primarily believes
in utilities nearest to }$u$ iff $\theta _{i}$'s level-1 belief only assigns
positive probability to $(c_{j},\theta _{j})$ which satisfies that $%
d(w_{j}(\theta _{j}),u_{j})\leq d(w_{j}(\theta _{j}^{\prime }),u_{j})$ for
all $\theta _{j}^{\prime }\in \Theta _{j}(\theta _{i})$ with $\beta
_{j}(\theta _{j}^{\prime })=\beta _{j}(\theta _{j}).$\smallskip \newline
\textbf{(3.2) }A type $\theta _{i}$ $\in \Theta _{i}$ has $u$\emph{-centered
belief} iff for any $c_{j},c_{j}^{\prime }\in C_{j},$ any $\theta _{j}\in
\Theta _{j}$, and any $v_{j},v_{j}^{\prime }\in V_{j}$ such that $%
(c_{j},\theta _{j}^{v_{j}})$ and $(c_{j}^{\prime },\theta
_{j}^{v_{j}^{\prime }})$ are deemed possible by $\theta _{i},$ it holds that 
$\theta _{i}$ deems $(c_{j},\theta _{j}^{v_{j}})$ infinitely more likely
than $(c_{j}^{\prime },\theta _{j}^{v_{j}^{\prime }})$ whenever $%
d(v_{j},u_{j})<d(v_{j}^{\prime },u_{j}).$\smallskip

Definition \ref{sec:asu}.3 gives restrictions on the order of types in a
lexicographic belief. (3.1) requires that $\theta _{i}$ primarily believes
in type $\theta _{j}$ only if $\theta _{j}$'s utility function is the
nearest one to $u_{j}$ among all types deemed possible by $\theta _{i}$
which share the same belief with $\theta _{j}.$ (3.2) requires that the
types of $j$ sharing the same belief deemed possible by $\theta _{i}$ are
arranged according to the distance of their assigned utility functions from $%
u_{j}:$ the farther a type $\theta _{j}$'s utility function is from $u_{j},$
the later $\theta _{j}$ occurs in the lexicographic belief of $\theta _{i}$.
(3.2) is a faithful translation of Perea and Roy (2017)'s Definition
3.2 into lexicographic model and (3.1) is weaker than (3.2).\smallskip

The essential difference between our conditions and Perea and Roy (2017)'s lies in the following definition.\smallskip \newline
\textbf{Definition \ref{sec:asu}.4 (A best (better) choice is supported by
utilities nearest (nearer) to }$u$\textbf{)}. Consider a static game form $%
G=(C_{i})_{i\in I},$ a\emph{\ }lexicographic epistemic model $M^{in}=(\Theta
_{i},w_{i},\beta _{i})_{i\in I}$ for\emph{\ }$G$\emph{\ }with incomplete
information, and a pair $u=(u_{i})_{i\in I}$ of utility functions.\smallskip 
\newline
\textbf{(4.1)} A type $\theta _{i}\in \Theta _{i}$ \emph{believes in that a
best choice of }$j$ \emph{is supported by utilities nearest to }$u$ iff for
any $(c_{j},\theta _{j}),$ $(c_{j}^{\prime },\theta _{j}^{\prime })$ deemed
possible by $\theta _{i}$ with $\beta _{j}(\theta _{j})=\beta _{j}(\theta
_{j}^{\prime })$, if $c_{j}$ is optimal for $\beta _{j}(\theta _{j})$ in $%
u_{j}$ but $c_{j}^{\prime }$ is not, then $d(w_{j}(\theta
_{j}),u_{j})<d(w_{j}(\theta _{j}^{\prime }),u_{j}).$\smallskip \newline
\textbf{(4.2)} A type $\theta _{i}\in \Theta _{i}$ \emph{believes in that a
better choice of }$j$ \emph{is supported by utilities nearer to }$u$ iff for
any $(c_{j},\theta _{j}),$ $(c_{j}^{\prime },\theta _{j}^{\prime })$ deemed
possible by $\theta _{i}$ with $\beta _{j}(\theta _{j})=\beta _{j}(\theta
_{j}^{\prime }),$ if $u_{j}(c_{j},\theta _{j})>u_{j}(c_{j}^{\prime },\theta
_{j}^{\prime }),$ then $d(w_{j}(\theta _{j}),u_{j})<d(w_{j}(\theta
_{j}^{\prime }),u_{j}).$\smallskip

Definition \ref{sec:asu}.4 gives restriction on the relation between paired
choices and types. (4.1) requires that for each belief of player $j,$ a
choice optimal for that belief should be supported by the nearest utility
function to $u_{j}.$ (4.2) requires that for each belief of player $j$, a
utility function supporting a \textquotedblleft better\textquotedblright\
choice (i.e., $c_{j}$) should be nearer to $u_{j}$ than one supporting a
\textquotedblleft worse\textquotedblright\ choice (i.e., $c_{j}^{\prime }$).
It can be seen that (4.2) is stronger than (4.1).

(4.2) is similar to Perea and Roy (2017)'s Definition 3.3 which
requires that for each $(c_{j},\theta _{j})$ deemed possible by $\theta
_{i}, $ $w_{j}(\theta _{j})$ is the nearest utility function in $V_{j}$ to $%
u_{j}$ among those in which $c_{j}$ is rational under $\beta _{j}(\theta
_{j})$. It can be shown by Lemmas 5.4 and 5.5 in Perea and Roy (2017)
that Definition \ref{sec:asu}.4 (4.2) is weaker than Perea and Roy \cite%
{ps17}'s Definition 3.3. We adopt it here since a nearest utility function
does not in general exist for lexicographic beliefs. That is, given $%
u_{j}\in V_{j},$ $c_{j}\in C_{j}$, and a lexicographic belief $\beta _{j},$
there may not exist $v_{j}\in V_{j}$ satisfying that (1) $c_{j}$ is rational
at $v_{j}$ under $\beta _{j},$ and (2) there is no $v_{j}^{\prime }\in V_{j}$
such that $c_{j}$ is rational at $v_{j}^{\prime }$ for $\beta _{j}$ and $%
d(v_{j}^{\prime },u_{j})<d(v_{j},u_{j}).$ See the following
example.\smallskip \newline
\textbf{Example \ref{sec:asu}.1 (No nearest utility function)}. Consider a
game $\Gamma $ where player $1$ has choices $A,B,$ and $C$ and player $2$
has choices $D,E,$ and $F.$ The payoff function $u_{1}$ of player $1$ is shown in Table 5.
\begin{table}
\label{tab:5}
\caption{No nearest utility function in lexicographic beliefs}
\begin{tabular}{llll}
\hline \noalign{\smallskip}
$u_{1}$ & $D$ & $E$ & $F$ \\ 
\noalign{\smallskip} \hline \noalign{\smallskip}
$A$ & $1$ & $1$ & $1$ \\ 
$B$ & $1$ & $1$ & $0$\\
$C$ & $1$ & $0$ & $1$\\
\noalign{\smallskip} \hline
\end{tabular}%
\end{table}
Let $\beta _{1}=(D,E,F),$ that is, player $1$ deems player $2$'s choice $D$
infinitely more likely than $E$ and $E$ infinitely more likely than $F$. In $%
u_{1},$ $A$ is rational for $\beta _{1}$ but $B$ is not. 

Now we show that
there is no nearest utility function to $u_{1}$ at which $B$ is rational
under $\beta _{1}$. Suppose there is such a function $v_{1}\in V_{1}.$ Let $%
d=d(v_{1},u_{1}).$ It can be seen that $d>0.$ Consider $v_{1}^{\prime }$ in Table 6.
\begin{table}
\label{tab:6}
\caption{A worse choice is supported by a better utility function}
\begin{tabular}{llll}
\hline \noalign{\smallskip}
$u_{1}$ & $D$ & $E$ & $F$ \\ 
\noalign{\smallskip} \hline \noalign{\smallskip}
$A$ & $1$ & $1$ & $1$ \\ 
$B$ & $1+\frac{d}{2}$ & $1$ & $0$\\
$C$ & $1$ & $0$ & $1$\\
\noalign{\smallskip} \hline
\end{tabular}%
\end{table}
$B$ is also rational at $v_{1}^{\prime }$ under $\beta _{1},$ while $%
d(v_{1}^{\prime },u_{1})=\frac{d}{2}<d=d(v_{1},u_{1}),$ a contradiction.
Also, though $\beta _{1}$ prefers $B$ to $C$ in $u_{1},$ it can be seen that
for each utility function $v_{1}^{B}$ in which $B$ is rational under $\beta
_{1}$, there is some $v_{1}^{C}\in V_{1}$ satisfying (1) $C$ is optimal in $%
v_{1}^{C}$ under $\beta _{1},$ and (2) $%
d(v_{1}^{C},u_{1})<d(v_{1}^{B},u_{1}).$ Indeed, this can be done by letting $%
v_{1}^{C}(C,D)=1+d(v_{1}^{B},u_{1})/2$ and $%
v_{1}^{C}(c_{1},c_{2})=u_{1}(c_{1},c_{2})$ for all other $(c_{1},c_{2})\in
C_{1}\times C_{2}$.\smallskip

Example \ref{sec:asu}.1 shows that the relationship between preferences
among choices and the distance of utility functions from the original one is
more complicated for lexicographic beliefs. That is why we adopt Definition %
\ref{sec:asu}.4 (4.2) here. The following lemma guarantees the existence of
utility functions satisfying the condition in Definition \ref{sec:asu}.4
(4.2). It shows that, given a utility function $u_{i}$ and a lexicographic
belief $\beta _{i},$ corresponding to the sequence $c_{i1},...,c_{iN}$ of $i$%
's choices arranged from the most to the least preferred at $u_{i}$ under $%
\beta _{i}$, there is a sequence $v_{i1},...,v_{iN}$ of utility functions
arranged from the nearest to the farthest to $u_{i}$ such that for each $%
n=1,...,N,$ $c_{in}$ is rational at $v_{in}$ under $\beta _{i}.$ This lemma
plays a similar role in our characterizations as Lemmas 5.4 and 5.5 in Perea
and Roy \cite{ps17}.\smallskip \newline
\textbf{Lemma \ref{sec:asu}.1 (Existence of utilities satisfying Definition %
\ref{sec:asu}.4 (4.2))}. Consider a static game form $G=(C_{i})_{i\in I},$ $%
u_{i}\in V_{i},$ and $\beta _{i}=(\beta _{i1},\beta _{i2},...,\beta _{iK})$
such that $\beta _{ik}\in \Delta (C_{j})$ for each $k=1,...,K.$ Let $\Pi
_{i}(\beta _{i})=(C_{i1},C_{i2},...,C_{iL})$ be a partition of $C_{i}$
satisfying that (1) for each $\ell =1,...,L$ and each $c_{i\ell },c_{i\ell
}^{\prime }\in C_{i\ell },$ $u_{i}(c_{i\ell },\beta _{i})=u_{i}(c_{i\ell
}^{\prime },\beta _{i}),$ and (2) for each $\ell =1,...,L-1,$ each $c_{i\ell
}\in C_{i\ell }$ and $c_{i,\ell +1}\in C_{i,\ell +1},$ $u_{i}(c_{i\ell
},\beta _{i})>u_{i}(c_{i,\ell +1},\beta _{i})$. That is, $\Pi _{i}(\beta
_{i})$ is the sequence of equivalence classes of choices in $C_{i}$ arranged
from the most preferred to the least preferred under $\beta _{i}.$

Then there are $v_{i1},...,v_{iL}\in V_{i}$ satisfying\smallskip \newline
(a) $v_{i1}=u_{i},$\smallskip \newline
(b) For each $\ell =1,...,L$ and each $c_{i\ell }\in C_{i\ell },$ $c_{i\ell
} $ is rational at $v_{i\ell }$ under $\beta _{i},$ and\smallskip \newline
(c) For each $\ell =1,...,L-1,$ $d(v_{i\ell },u_{i})<d(v_{i,\ell +1},u_{i}).$

\subsection{The second group of conditions}

\textbf{Definition \ref{sec:asu}.5 (Belief in rationality)}. $\theta _{i}\in
\Theta _{i}$ \emph{believes in }$j$\emph{'s rationality }iff $\theta _{i}$
deems $(c_{j},\theta _{j})$ possible only if $c_{j}$ is rational for $\theta
_{j}.$\smallskip

In an incomplete information model, since each type is assigned with a
belief on the opponent's choice-type pairs as well as a payoff function,
caution and a full belief of rationality can be satisfied simultaneously.
The consistency of caution and rationality is the essential difference of
models with incomplete information from those with complete information.
Rationality does not appear in the first group of characterizations, but in
the proofs we will construct incomplete information models whose types
satisfies all the conditions as well as common full belief in rationality.
On the other hand, rationality plays an important role in the
characterization of the second group. We will discuss more about this
consistency between caution and rationality in Sections \ref{sec:cr}%
.3.\smallskip \newline
\textbf{Definition \ref{sec:asu}.6 (Primary belief in }$u$ \textbf{and prior
assumption of }$u$\textbf{)}. Consider a static game form $G=(C_{i})_{i\in
I},$ a\emph{\ }lexicographic epistemic model $M^{in}=(\Theta
_{i},w_{i},\beta _{i})_{i\in I}$ for\emph{\ }$G$\emph{\ }with incomplete
information, and a pair $u=(u_{i})_{i\in I}$ of utility functions.\smallskip 
\newline
\textbf{(6.1)} $\theta _{i}\in \Theta _{i}$ \emph{primarily believes in }$u$
iff $\theta _{i}$'s level-1 belief only assigns positive probability to $%
(c_{j},\theta _{j})$ with $w_{j}(\theta _{j})=u_{j}.$\smallskip \newline
\textbf{(6.2)} $\theta _{i}\in \Theta _{i}$ \emph{prior assumes }$u$ iff for
any $(c_{j},\theta _{j})$ with $\theta _{j}$ cautious deemed possible by $%
\theta _{i}$ satisfying that $w_{j}(\theta _{j})=u_{j},$ then $\theta _{i}$
deems $(c_{j},\theta _{j})$ infinitely more likely than any pair does not
satisfy that property.\smallskip

Primary belief in $u$ is stronger than Definition \ref{sec:asu}.3 (3.1).
(3.1) allows the occurrence of a type with a utility function which is very
similar\ (but not necessarily equal) to $u_{j}$ in the level-1 belief of $%
\theta _{i},$ while primary belief in $u$ only allows types with utility
function $u_{j}$ there. (6.2) is stronger than (6.1) since (6.2) requires
that all choice-type pairs which believes in $u$ should be put in front of
pairs which do not believe in $u,$ while (6.1) only requires that level-1
belief believes in $u.$\smallskip \newline
\textbf{Definition \ref{sec:asu}.7 (Every good choice is supported)}.
Consider a static game form $G=(C_{i})_{i\in I},$ a\emph{\ }lexicographic
epistemic model $M^{in}=(\Theta _{i},w_{i},\beta _{i})_{i\in I}$ for\emph{\ }%
$G$\emph{\ }with incomplete information, and a pair $u=(u_{i})_{i\in I}$ of
utility functions. A cautious type $\theta _{i}\in \Theta _{i}$ \emph{%
assumes that every good choice of }$j$ \emph{is supported }iff for each $%
c_{j}$ that is optimal for some cautious type of $j$ (may not be in $M^{in}$%
) with $u_{j}$ as its assigned utility function, $\theta _{i}$ deems
possible a cautious type $\theta _{j}\in \Theta _{j}$ such that $%
w_{j}(\theta _{j})=u_{j}$ and $c_{j}$ is optimal for $\theta _{j}.$\smallskip

Common assumption of prior $u$ and that every good choice is supported is,
as common assumption of rationality in incomplete information model, more
complicated from common full belief. We have the following
definition.\smallskip \newline
\textbf{Definition \ref{sec:asu}.8 (}$n$\textbf{-fold} \textbf{assumption of
prior }$u$\textbf{\ and} \textbf{that every good choice is supported) }%
Consider a static game form $G=(C_{i})_{i\in I},$ a\emph{\ }lexicographic
epistemic model $M^{in}=(\Theta _{i},w_{i},\beta _{i})_{i\in I}$ for\emph{\ }%
$G$\emph{\ }with incomplete information, and a pair $u=(u_{i})_{i\in I}$ of
utility functions. $\theta _{i}\in \Theta _{i}$ \emph{expresses }$1$\emph{%
-fold assumption of prior }$u$ \emph{and} \emph{that every good choice is
supported\ }iff it prior assumes $u$ and assumes that every good choice of $%
j $ is supported. For any $n\in \mathbb{N},$ we say that a cautious type $%
\theta _{i}\in \Theta _{i}$ expresses $(n+1)$\emph{-fold assumption of prior 
}$u$ \emph{and that every good choice is supported} iff the following two
conditions are satisfied:\smallskip \newline
(a) whenever a choice $c_{j}$ of player $j$ is optimal for some cautious
type (not necessarily in $M^{in}$) with $u_{j}$ as its assigned utility
function that expresses up to $n$-fold assumption of that every good choice
is supported, $\theta _{i}$ deems possible some cautious type $\theta _{j}$
with $w_{j}(\theta _{j})=u_{j}$ for player $j$ which expresses up to $n$%
-fold assumption of prior $u$ and that every good choice is supported for
which $c_{j}$ is optimal.\smallskip \newline
(b) $\theta _{i}$ deems all choice-type pairs $(c_{j},\theta _{j})$, where $%
\theta _{j}$ is cautious and expresses up to $n$-fold assumption of prior $u$
and that every good choice is supported and satisfies $w_{j}(\theta
_{j})=u_{j},$ infinitely more likely than any choice-type pair $%
(c_{j}^{\prime },\theta _{j}^{\prime })$ that does not satisfy this
property.\smallskip

We say that $t_{i}$ \emph{expresses common assumption of prior }$u$ \emph{%
and in that every good choice is supported} iff it expresses $n$-fold
assumption of prior $u$ and that every good choice is supported for every $%
n\in \mathbb{N}.$

\section{Characterizations}
\label{sec:cce}

So far we have introduced conditions under two different frameworks for
static games: one includes permissibility, proper rationalizability, and
iterated admissibility within a complete information framework, the other
contains various conditions on types within an incomplete information
framework. In this section we will show that there are correspondences
between them. The characterizations will be separated into two groups: in
the first group we characterize permissibility and proper rationalizability,
and in the second group we characterize permissibility in an alternative way
and iterated admissibility. The critical difference between the two groups
is the role of rationality.

\subsection{Without rationality: permissibility and proper rationalizability}

In this subsection we give characterizations of permissibility and proper
rationalizability. An illustrative example will also be provided.\smallskip 
\newline
\textbf{Theorem \ref{sec:cce}.1 (Characterization of permissibility)}.
Consider a finite 2-person static game $\Gamma =(C_{i},u_{i})_{i\in I}$ and
the corresponding game form $G=(C_{i})_{i\in I}.$

Then, $c_{i}^{\ast }\in C_{i}$ is permissible if and only if there is some
finite lexicographic epistemic model $M^{in}=(\Theta _{i},w_{i},\beta
_{i})_{i\in I}$ with incomplete information for $G$ and some $\theta
_{i}^{\ast }\in \Theta _{i}$ with $w_{i}(\theta _{i}^{\ast })=u_{i}$ such
that\smallskip \newline
(a) $c_{i}^{\ast }$ is rational for $\theta _{i}^{\ast },$ and,\smallskip 
\newline
(b) $\theta _{i}^{\ast }$ expresses common full belief in caution, primary
belief in utilities nearest to $u$, and that a best choice is supported by
utilities nearest to $u$.\smallskip \newline
\textbf{Theorem \ref{sec:cce}.2 (Characterization of proper
rationalizability)}. Consider a finite 2-person static game $\Gamma
=(C_{i},u_{i})_{i\in I}$ and the corresponding game form $G=(C_{i})_{i\in
I}. $

Then, $c_{i}^{\ast }\in C_{i}$ is properly rationalizable if and only if
there is some finite lexicographic epistemic model $M^{in}=(\Theta
_{i},w_{i},\beta _{i})_{i\in I}$ for $G$ and some $\theta _{i}^{\ast }\in
\Theta _{i}$ with $w_{i}(\theta _{i}^{\ast })=u_{i}$ such that\smallskip 
\newline
(a) $c_{i}^{\ast }$ is rational for $\theta _{i}^{\ast }$, and\smallskip 
\newline
(b) $\theta _{i}^{\ast }$ expresses common full belief in caution, $u$%
-centered belief, and that a better choice is supported by utilities nearer
to $u$.\smallskip

To show these statements, we will construct a correspondence between
complete information models and incomplete ones and show that conditions on
a type in one model can be transformed into proper conditions on the
corresponding type in the constructed model. We use the following example to
show the intuition.\smallskip \newline
\textbf{Example \ref{sec:cce}.1}. Consider the game $\Gamma $
in Table 7(Perea, 2012, p.190) and the lexicographic model $M^{co}=(T_{i},b_{i})_{i\in I}$ for $\Gamma $
where $T_{1}=\{t_{1}\},$ $T_{2}=\{t_{2}\}$, and%
\begin{equation*}
b_{1}(t_{1})=((D,t_{2}),(E,t_{2}),(F,t_{2})),\text{ }%
b_{2}(t_{2})=((C,t_{1}),(B,t_{1}),(A,t_{1})).
\end{equation*}%
\begin{table}
\label{tab:7}
\caption{The game for Example \ref{sec:cce}.1}
\begin{tabular}{llll}
\hline \noalign{\smallskip}
$u_{1} \backslash u_{2}$ & $D$ & $E$ & $F$ \\ 
\noalign{\smallskip} \hline \noalign{\smallskip}
$A$ & $0,3$ & $1,2$ & $1,1$ \\ 
$B$ & $1,3$ & $0,2$ & $1,1$\\
$C$ & $1,3$ & $1,2$ & $0,1$\\
\noalign{\smallskip} \hline
\end{tabular}%
\end{table}
It can be seen that $D$ is properly rationalizable (and therefore
permissible) since it is rational for $t_{2}$ which expresses common full
belief in caution and respect of preferences. Consider the lexicographic
epistemic model $M^{in}=(\Theta _{i},w_{i},\beta _{i})_{i\in I}$ with
incomplete information for the corresponding game form where $\Theta
_{1}=\{\theta _{11},\theta _{12},\theta _{13}\},$ $\Theta _{2}=\{\theta
_{21},\theta _{22},\theta _{23}\},$ and%
\begin{eqnarray*}
w_{1}(\theta _{11}) &=&u_{1},\text{ }\beta _{1}(\theta _{11})=((D,\theta
_{21}),(E,\theta _{22}),(F,\theta _{23})), \\
w_{1}(\theta _{12}) &=&v_{1},\text{ }\beta _{1}(\theta _{12})=((D,\theta
_{21}),(E,\theta _{22}),(F,\theta _{23})), \\
w_{1}(\theta _{13}) &=&v_{1}^{\prime },\text{ }\beta _{1}(\theta
_{13})=((D,\theta _{21}),(E,\theta _{22}),(F,\theta _{23})), \\
w_{2}(\theta _{21}) &=&u_{2},\text{ }\beta _{2}(\theta _{21})=((C,\theta
_{11}),(B,\theta _{12})),(A,\theta _{13})), \\
w_{2}(\theta _{22}) &=&v_{2},\text{ }\beta _{2}(\theta _{22})=((C,\theta
_{11}),(B,\theta _{12})),(A,\theta _{13})), \\
w_{2}(\theta _{23}) &=&v_{2}^{\prime },\text{ }\beta _{2}(\theta
_{23})=((C,\theta _{11}),(B,\theta _{12})),(A,\theta _{13})).
\end{eqnarray*}%
where the alternative utility functions are shown in Table 8.
\begin{table}
\label{tab:8}
\caption{Alternative utility functions for players $1$ and $2$}
\begin{tabular}{llll}
\hline \noalign{\smallskip}
$v_{1}$ & $D$ & $E$ & $F$ \\ 
\noalign{\smallskip} \hline \noalign{\smallskip}
$A$ & $0$ & $1$ & $1$ \\ 
$B$ & $2$ & $0$ & $1$\\ 
$C$ & $1$ & $1$ & $0$\\
\noalign{\smallskip} \hline
\end{tabular}\quad 
\begin{tabular}{llll}
\hline \noalign{\smallskip}
$v_{1}^{\prime}$ & $D$ & $E$ & $F$ \\ 
\noalign{\smallskip} \hline \noalign{\smallskip}
$A$ & $3$ & $1$ & $1$ \\ 
$B$ & $2$ & $0$ & $1$\\ 
$C$ & $1$ & $1$ & $0$\\
\noalign{\smallskip} \hline
\end{tabular}\quad
\begin{tabular}{llll}
\hline \noalign{\smallskip}
$v_{2}$ & $D$ & $E$ & $F$ \\ 
\noalign{\smallskip} \hline \noalign{\smallskip}
$A$ & $3$ & $2$ & $1$ \\ 
$B$ & $3$ & $2$ & $1$\\ 
$C$ & $3$ & $4$ & $1$\\
\noalign{\smallskip} \hline
\end{tabular}\quad 
\begin{tabular}{llll}
\hline \noalign{\smallskip}
$v_{2}^{\prime}$ & $D$ & $E$ & $F$ \\ 
\noalign{\smallskip} \hline \noalign{\smallskip}
$A$ & $3$ & $2$ & $1$ \\ 
$B$ & $3$ & $2$ & $1$\\ 
$C$ & $3$ & $4$ & $5$\\
\noalign{\smallskip} \hline
\end{tabular}
\end{table}
For each $i\in I,$ $\theta _{i1}$, $\theta _{i2},$ and $\theta _{i3}$ have
the same belief; the only difference lies in their assigned utility
functions since each should support some choice. The relation between $%
M^{in} $ and $M^{co}$ can be seen clearly: for each $i\in I,$ $\theta _{i1}$%
, $\theta _{i2},$ and $\theta _{i3}$ correspond to $t_{i}$ in the sense that
the belief of the former is obtained by replacing every occurrence of $t_{j}$
in the belief of $t_{i}$ by the type corresponding to $t_{j}$ in $M^{in}$ at
which the paired choice is optimal. It can be seen that $\theta _{11}$
expresses common full belief in caution, $u$-centered belief, and that a
better choice is supported by utilities nearer to $u$ (therefore primary
belief in utilities nearest to $u$ and that a best choice is supported by
utilities nearest to $u$). Also, since the assigned utility function of $%
\theta _{11}$ is $u_{1},$ $C$ is rational for $\theta _{11}.$

\subsection{With rationality: permissibility and iterated admissibility}

In this subsection we give an alternative characterization of permissibility
and a characterization of iterated admissibility. Since the correspondence
between models with complete information and incomplete information are
basically the same with the previous group, here the results are given
without examples.\smallskip \newline
\textbf{Theorem \ref{sec:cce}.3 (An alternative characterization of
permissibility)}. Consider a finite 2-person static game $\Gamma
=(C_{i},u_{i})_{i\in I}$ and the corresponding game form $G=(C_{i})_{i\in I}$%
.

Then, $c_{i}^{\ast }\in C_{i}$ is permissible in $M^{co}$ if and only if
there is some finite lexicographic epistemic model $M^{in}=(\Theta
_{i},w_{i},\beta _{i})_{i\in I}$ with incomplete information for $G$ and
some $\theta _{i}^{\ast }\in \Theta _{i}$ with $w_{i}(\theta _{i}^{\ast
})=u_{i}$ such that\smallskip \newline
(a) $c_{i}^{\ast }$ is rational for $\theta _{i}^{\ast },$ and,\smallskip 
\newline
(b) $\theta _{i}^{\ast }$ expresses common full belief in caution,
rationality, and primary belief in $u$.\smallskip \newline
\textbf{Theorem \ref{sec:cce}.4 (Characterization of iterated admissibility)}%
. Consider a finite 2-person static game $\Gamma =(C_{i},u_{i})_{i\in I}$
and the corresponding game form $G=(C_{i}).$

Then $c_{i}^{\ast }\in C_{i}$ is iteratively admissible if and only if there
is some finite epistemic model $M^{in}=(\Theta _{i},w_{i},\beta _{i})_{i\in
I}$ with incomplete information for $G$ and some $\theta _{i}^{\ast }\in
\Theta _{i}$ with $w_{i}(\theta _{i}^{\ast })=u_{i}$ such that\smallskip 
\newline
(a) $c_{i}^{\ast }$ is rational for $\theta _{i}^{\ast }$, and\smallskip 
\newline
(b) $\theta _{i}^{\ast }$ expresses common full belief in caution,
rationality, and common assumption of prior $u$ and that every good choice
is supported.

\subsection{Proof of Theorem \protect\ref{sec:cce}.1}

To show the only-if part of Theorem \ref{sec:cce}.1 (and that of all other
theorems), we construct the following mapping from finite lexicographic
epistemic models with complete information to those with incomplete
information. Consider $\Gamma =(C_{i},u_{i})_{i\in I}$ and a finite
lexicographic epistemic model $M^{co}=(T_{i},b_{i})_{i\in I}$ with complete
information for $\Gamma .$ We first define types in a model with incomplete
information in the following two steps:\smallskip \newline
\textbf{Step 1}. For each $i\in I$ and $t_{i}\in T_{i},$ let $\Pi
_{i}(t_{i})=(C_{i1},...,C_{iL})$ be the sequence of equivalence classes of
choices in $C_{i}$ arranged from the most preferred to the least preferred
under $t_{i}.$ By Lemma \ref{sec:asu}.1, for each $C_{i\ell }$ there is some 
$v_{i\ell }(t_{i})\in V_{i}$ such that each choice in $C_{i\ell }$ is
rational at $v_{i\ell }(t_{i})$ under $t_{i},$ and $%
0=d(v_{i1}(t_{i}),u_{i})<d(v_{i2}(t_{i}),u_{i})<...<d(v_{iL}(t_{i}),u_{i}).$%
\smallskip \newline
\textbf{Step 2}. We define $\Theta _{i}(t_{i})=\{\theta
_{i1}(t_{i}),...,\theta _{iL}(t_{i})\}$ where for each $\ell =1,...,L,$ the
type $\theta _{i\ell }(t_{i})$ satisfies that (1) $w_{i}(\theta _{i\ell
}(t_{i}))=v_{i\ell }(t_{i}),$ and (2) $\beta _{i}(\theta _{i\ell }(t_{i}))$
is obtained from $b_{i}(t_{i})$ by replacing every $(c_{j},t_{j})$ with $%
c_{j}\in C_{jr}\in \Pi _{j}(t_{j})$ for some $r$ with $(c_{j},\theta _{j})$
where $\theta _{j}=\theta _{jr}(t_{j}),$ that is, $w_{j}(\theta _{j})$ is
the utility function among those corresponding to $\Pi _{j}(t_{j})$ in which 
$c_{j}$ is the rational for $t_{i}.$\smallskip

For each $i\in I,$ let $\Theta _{i}=\cup _{t_{i}\in T_{i}}\Theta
_{i}(t_{i}). $ Here we have constructed a finite lexicographic epistemic
model $M^{in}=(\Theta _{i},w_{i},\beta _{i})_{i\in I}$ for the corresponding
game form $G=(C_{i})_{i\in I}$ with incomplete information. It should be
noted that $M^{in}$ is not uniquely determined since there are multiple
sequences of utility functions satisfying the conditions in Lemma \ref%
{sec:asu}.1.

We show how this construction works by the following example.\smallskip 
\newline
\textbf{Example \ref{sec:cce}.2}. Consider the game $\Gamma $
in Table 9 (Perea, 2012, p.188) and the lexicographic epistemic model $M^{co}=(T_{i},b_{i})_{i\in I}$ of $%
\Gamma $ where $T_{1}=\{t_{1}\},$ $T_{2}=\{t_{2}\}$, and%
\begin{equation*}
b_{1}(t_{1})=((D,t_{2}),(C,t_{2})),\text{ }%
b_{2}(t_{2})=((A,t_{1}),(B,t_{1})).
\end{equation*}
\begin{table}
\label{tab:9}
\caption{The game for Example \ref{sec:cce}.2}
\begin{tabular}{lll}
\hline \noalign{\smallskip}
$u_{1}\backslash u_{2}$ & $C$ & $D$ \\
\noalign{\smallskip} \hline \noalign{\smallskip}
$A$ & $1,0$ & $0,1$ \\
$B$ & $0,0$ & $0,1$ \\
\noalign{\smallskip} \hline
\end{tabular}
\end{table}
We show how to construct a corresponding model $M^{in}=(\Theta
_{i},w_{i},\beta _{i})_{i\in I}$. First, by Step 1 it can be seen that $\Pi
_{1}(t_{1})=(\{A\},\{B\})$ and $\Pi _{2}(t_{2})=(\{D\},\{C\}).$ We let $%
v_{11}(t_{1})=u_{1}$ where $A$ is rational for $t_{1}$ and $v_{12}(t_{1})$
where $B$ is rational for $t_{1}$ as follows. Similarly, we let $%
v_{21}(t_{2})=u_{2}$ where $D$ is rational under $t_{2}$ and $v_{22}(t_{2})$
where $C$ is rational under $t_{2}$ as follows:%
\begin{table}
\label{tab:10}
\caption{Utility functions of player $1$ corresponding to $t_{1}$ and $t_{2}$}
\begin{tabular}{lll}
\hline \noalign{\smallskip}
$v_{12}(t_{1})$ & $C$ & $D$ \\
\noalign{\smallskip} \hline \noalign{\smallskip}
$A$ & $1$ & $0$ \\ 
$B$ & $0$ & $1$ \\
\noalign{\smallskip} \hline
\end{tabular}\quad
\begin{tabular}{lll}
\hline \noalign{\smallskip}
$v_{22}(t_{2})$ & $C$ & $D$ \\
\noalign{\smallskip} \hline \noalign{\smallskip}
$A$ & $2$ & $1$ \\ 
$B$ & $0$ & $1$ \\ 
\noalign{\smallskip} \hline
\end{tabular}%
\end{table}%

Then we go to Step 2. It can be seen that $\Theta _{1}(t_{1})=\{\theta
_{11}(t_{1}),\theta _{12}(t_{1})\},$ where%
\begin{eqnarray*}
w_{1}(\theta _{11}(t_{1})) &=&v_{11}(t_{1}),\text{ }\beta _{1}(\theta
_{11}(t_{1}))=((D,\theta _{21}(t_{2})),(C,\theta _{22}(t_{2}))), \\
w_{1}(\theta _{12}(t_{1})) &=&v_{12}(t_{1}),\text{ }\beta _{1}(\theta
_{12}(t_{1}))=((D,\theta _{21}(t_{2})),(C,\theta _{22}(t_{2}))).
\end{eqnarray*}%
Also, $\Theta _{2}(t_{2})=\{\theta _{21}(t_{2}),\theta _{22}(t_{2})\},$ where%
\begin{eqnarray*}
w_{2}(\theta _{21}(t_{2})) &=&v_{21}(t_{2}),\text{ }\beta _{2}(\theta
_{21}(t_{2}))=((A,\theta _{11}(t_{1})),(B,\theta _{12}(t_{1}))), \\
w_{2}(\theta _{22}(t_{2})) &=&v_{22}(t_{2}),\text{ }\beta _{2}(\theta
_{22}(t_{2}))=((A,\theta _{11}(t_{1})),(B,\theta _{12}(t_{1}))).
\end{eqnarray*}

Consider $M^{co}=(T_{i},b_{i})_{i\in I}$ and $M^{in}=(\Theta
_{i},w_{i},\beta _{i})_{i\in I}$ corresponding to $M^{co}$ constructed by
the two steps above. We have the following observations.\smallskip \newline
\textbf{Observation \ref{sec:cce}.1 (Redundancy)}. For each $t_{i}\in T_{i}$
and each $\theta _{i},\theta _{i}^{\prime }\in \Theta _{i}(t_{i}),$ $\beta
_{i}(\theta _{i})=\beta _{i}(\theta _{i}^{\prime }).$\smallskip \newline
\textbf{Observation \ref{sec:cce}.2 (Rationality)}. Each $\theta _{i}\in
\Theta _{i}(t_{i})$ believes in $j$'s rationality.\smallskip \newline
\textbf{Observation \ref{sec:cce}.3 (A better choice is supported by
utilities nearer to }$u$\textbf{)}. Each $\theta _{i}\in \Theta _{i}(t_{i})$
believes that a better choice is supported by utilities nearer to $u$%
.\smallskip

The observations are true by construction. Observation \ref{sec:cce}.1 means
that the difference between any two types in a $\Theta _{i}(t_{i})$ is in
the utility functions assigned to them. Observation \ref{sec:cce}.2 means
that in an incomplete information model constructed from one with complete
information, each type believes in the opponent's rationality. This is
because in the construction, we requires that for each pair $(c_{j},t_{j})$
occurring in a belief, its counterpart in the incomplete information
replaces $t_{j}$ by type in $\Theta _{j}(t_{j})$ with the utility function
in which $c_{j}$ is optimal for $t_{j}$. It follows from Observation \ref%
{sec:cce}.2 that each $\theta _{i}\in \Theta _{i}(t_{i})$ expresses common
full belief in rationality. Observation \ref{sec:cce}.3 implies that the
best choice is supported by utilities nearest to $u.$ It follows that each $%
\theta _{i}\in \Theta _{i}(t_{i})$ expresses common full belief in that a
best (better) choice is supported by utilities nearest (nearer) to $u$%
.\smallskip

By construction, each $t_{i}$ shares the same belief about $j$'s choices at
each level with each $\theta _{i}\in \Theta _{i}(t_{i})$; also, for each $%
t_{i}\in T_{i},$ the utility function assigned to $\theta _{i1}(t_{i})$ is $%
u_{i}.$ It is clear that any $c_{i}$ rational for $t_{i}$ is also rational
for $\theta _{i1}(t_{i}).$ Therefore, to show the only-if part of Theorem %
\ref{sec:cce}.1, we show that if $t_{i}$ expresses common full belief in
caution and primary belief in rationality, then $\theta _{i1}(t_{i})$
expresses common belief in caution, primary belief in utilities nearest to $%
u $, and that a best choice is supported by utilities nearest to $u$%
.\smallskip \newline
\textbf{Lemma \ref{sec:cce}.1 (Caution}$^{co}\rightarrow $\textbf{\ Caution}$%
^{in}$\textbf{)}. Consider $M^{co}=(T_{i},b_{i})_{i\in I}$ and $%
M^{in}=(\Theta _{i},w_{i},\beta _{i})_{i\in I}$ corresponding to $M^{co}$.
If $t_{i}\in T_{i}$ expresses common full belief in caution, so does each $%
\theta _{i}\in \Theta _{i}(t_{i}).$\smallskip \newline
\textbf{Lemma \ref{sec:cce}.2 (Primary belief in rationality }$\rightarrow $%
\textbf{\ primary belief in utilities nearest to }$u$\textbf{)}. Consider $%
M^{co}=(T_{i},b_{i})_{i\in I}$ and $M^{in}=(\Theta _{i},w_{i},\beta
_{i})_{i\in I}$ corresponding to $M^{co}$. If $t_{i}\in T_{i}$ expresses
common full belief in primary belief in rationality, then each $\theta
_{i}\in \Theta _{i}(t_{i})$ expresses common full belief in primary belief
in utilities nearest to $u$.\smallskip \newline
\textbf{Proof of the only-if part of Theorem \ref{sec:cce}.1}. Consider $%
M^{co}=(T_{i},b_{i})_{i\in I}$, $M^{in}=(\Theta _{i},w_{i},\beta _{i})_{i\in
I}$ corresponding to $M^{co}$, a permissible choice $c_{i}^{\ast }\in C_{i}$%
, and $t_{i}^{\ast }\in T_{i}$ which is a type expressing common full belief
in caution and primary belief in rationality such that $c_{i}^{\ast }$ is
rational for $t_{i}^{\ast }.$ Let $\theta _{i}^{\ast }=\theta
_{i1}(t_{i}^{\ast }).$ By definition, $w_{i}(\theta _{i}^{\ast })=u_{i}$ and 
$\beta _{i}(\theta _{i}^{\ast })$ has the same distribution on $j$'s choices
at each level as $b_{i}(t_{i}^{\ast })$. Hence $c_{i}^{\ast }$ is rational
for $\theta _{i}^{\ast }.$ Also, it follows from Observation \ref{sec:cce}.3
and Lemmas \ref{sec:cce}.1 and \ref{sec:cce}.2 that $\theta _{i}^{\ast }$
expresses common full belief in caution, primary belief in utilities nearest
to $u$, and that a best choice is supported by utilities nearest to $u$. $%
\square $\smallskip

To show the if part, we need a mapping from models with incomplete
information to those with complete information. Consider a finite 2-person
static game $\Gamma =(C_{i},u_{i})_{i\in I},$ the corresponding game form $%
G=(C_{i})_{i\in I},$ and a finite epistemic model $M^{in}=(\Theta
_{i},w_{i},\beta _{i})_{i\in I}$ for $G$ with incomplete information. We
construct a model $M^{co}=(T_{i},b_{i})_{i\in I}$ for $\Gamma $ with
complete information as follows. For each $\theta _{i}\in \Theta _{i},$ we
define $E_{i}(\theta _{i})=\{\theta _{i}^{\prime }\in \Theta _{i}:\beta
_{i}(\theta _{i}^{\prime })=\beta (\theta _{i})\}.$ In this way $\Theta _{i}$
is partitioned into several equivalence classes $\mathbb{E}%
_{i}=\{E_{i1},...,E_{iL}\}$ where for each $\ell =1,..,L,$ $E_{i\ell
}=E_{i}(\theta _{i})$ for some $\theta _{i}\in \Theta _{i}.$ To each $%
E_{i}\in \mathbb{E}_{i}$ we use $t_{i}(E_{i})$ to denote a type. We define $%
b_{i}(t_{i}(E_{i}))$ to be a lexicographic belief which is obtained from $%
\beta _{i}(\theta _{i})$ by replacing each occurrence of $(c_{j},\theta
_{j}) $ by $(c_{j},t_{j}(E_{j}(\theta _{j})));$ in other words, $%
b_{i}(t_{i}(E_{i}))$ has the same distribution on choices at each level as $%
\beta _{i}(\theta _{i})$ for each $\theta _{i}\in E_{i},$ while each $\theta
_{j}\in \Theta _{j}(\theta _{i})$ is replaced by $t_{j}(E_{j}(\theta _{j})).$
For each $i\in I,$ let $T_{i}=\{t_{i}(E_{i})\}_{E_{i}\in \mathbb{E}_{i}}.$
We have constructed from $M^{in}$ a finite epistemic model $%
M^{co}=(T_{i},b_{i})_{i\in I}$ with complete information for $\Gamma .$

We show how this construction works by the following example.\smallskip 
\newline
\textbf{Example \ref{sec:cce}.3}. Consider the game $\Gamma $ in Example \ref%
{sec:cce}.2 and the model $M^{in}=(\Theta _{i},w_{i},\beta _{i})_{i\in I}$
for the corresponding game form where $\Theta _{1}=\{\theta _{11},\theta
_{12}\},$ $\Theta _{2}=\{\theta _{21},\theta _{22}\},$ and%
\begin{eqnarray*}
w_{1}(\theta _{11}) &=&u_{1},\text{ }\beta _{1}(\theta _{11})=((D,\theta
_{21}),(C,\theta _{22})), \\
w_{1}(\theta _{12}) &=&v_{1},\text{ }\beta _{1}(\theta _{12})=((D,\theta
_{21}),(C,\theta _{22})), \\
w_{2}(\theta _{21}) &=&u_{2},\text{ }\beta _{2}(\theta _{21})=((A,\theta
_{11}),(B,\theta _{12})), \\
w_{2}(\theta _{22}) &=&v_{2},\text{ }\beta _{2}(\theta _{22})=((A,\theta
_{11}),(B,\theta _{12})).
\end{eqnarray*}%
where $v_{1}=v_{12}(t_{1})$ and $v_{2}=v_{22}(t_{2})$ in Example \ref%
{sec:cce}.2. It can be seen that $\mathbb{E}_{1}=\{\{\theta _{11},\theta
_{12}\}\}$ since $\beta _{1}(\theta _{11})=\beta _{1}(\theta _{12})$ and $%
\mathbb{E}_{2}=\{\{\theta _{21},\theta _{22}\}\}$ since $\beta _{2}(\theta
_{21})=\beta _{2}(\theta _{22}).$ Corresponding to those equivalence classes
we have $t_{1}(\{\theta _{11},\theta _{12}\})$ and $t_{2}(\{\theta
_{21},\theta _{22}\}),$ and

\begin{eqnarray*}
b_{1}(t_{1}(\{\theta _{11},\theta _{12}\})) &=&((D,t_{2}(\{\theta
_{21},\theta _{22}\})),(C,t_{2}(\{\theta _{21},\theta _{22}\}))), \\
b_{2}(t_{2}(\{\theta _{21},\theta _{22}\})) &=&((A,t_{1}(\{\theta
_{11},\theta _{12}\})),(B,t_{1}(\{\theta _{11},\theta _{12}\}))).
\end{eqnarray*}

It can be seen that this is the reversion of the previous construction if
each type has distinct belief on choices of the opposite. Indeed, let $%
M^{co}=(T_{i},b_{i})_{i\in I}$ satisfying that $b_{i}(t_{i})\neq
b_{i}(t_{i}^{\prime })$ for each $t_{i},t_{i}^{\prime }\in T_{i}$ with $%
t_{i}\neq t_{i}^{\prime }$, and $M^{in}=(\Theta _{i},w_{i},\beta _{i})_{i\in
I}$ be constructed from $M^{co}$ by the previous two steps. Then $\mathbb{E}%
_{i}=\{\Theta _{i}(t_{i})\}_{t_{i}\in T_{i}}$ and $t_{i}(\Theta
_{i}(t_{i}))=t_{i}$ for each $i\in I.$\smallskip

To show the if part of Theorem \ref{sec:cce}.1, we need the following
lemmas.\smallskip \newline
\textbf{Lemma \textbf{\ref{sec:cce}}.3} \textbf{(Caution}$^{in}\rightarrow $%
\textbf{\ Caution}$^{co}$\textbf{)}. Consider $M^{in}=(\Theta
_{i},w_{i},\beta _{i})_{i\in I}$ and $M^{co}=(T_{i},b_{i})_{i\in I}$
corresponding to $M^{in}$ constructed by the above approach. If $\theta
_{i}\in \Theta _{i}$ expresses common full belief in caution, so does $%
t_{i}(E_{i}(\theta _{i})).$\smallskip \newline
\textbf{Lemma \ref{sec:cce}.4} \textbf{(Caution}$^{in}$\textbf{\ + primary
belief in utilities nearest to }$u$\textbf{\ + a best choice is supported by
utilities nearest to }$u\rightarrow $\textbf{\ Primary belief in rationality)%
}. Consider $M^{in}=(\Theta _{i},w_{i},\beta _{i})_{i\in I}$ and $%
M^{co}=(T_{i},b_{i})_{i\in I}$ corresponding to $M^{in}$. If $\theta _{i}\in
\Theta _{i}$ expresses common full belief in caution, primary belief in
utilities nearest to $u$, and that a best choice is supported by utilities
nearest to $u$, then $t_{i}(E_{i}(\theta _{i}))$ expresses common full
belief in primary belief in rationality.\smallskip \newline
\textbf{Proof of the if part of Theorem \ref{sec:cce}.1}. Consider $%
M^{co}=(T_{i},b_{i})_{i\in I}$ corresponding to $M^{in}$ and $c_{i}^{\ast
}\in C_{i}$ rational for some $\theta _{i}^{\ast }$ with $w_{i}(\theta
_{i}^{\ast })=u_{i}$ which expresses common full belief in caution, primary
belief in utilities nearest to $u$, and that a best choice is supported by
utilities nearest to $u$. Consider $t_{i}(E_{i}(\theta _{i}^{\ast })).$
Since $w_{i}(\theta _{i}^{\ast })=u_{i}$ and $b_{i}(t_{i}(E_{i}(\theta
_{i}^{\ast })))$ has the same distribution on $j$'s choices at each level as 
$\beta _{i}(\theta _{i}^{\ast })$, $c_{i}^{\ast }$ is rational for $%
t_{i}(E_{i}(\theta _{i}^{\ast })).$ Also, by Lemmas \ref{sec:cce}.3 and \ref%
{sec:cce}.4, $t_{i}(E_{i}(\theta _{i}^{\ast }))$ expresses common full
belief in\ caution and primary belief in rationality. Hence $c_{i}^{\ast }$
is permissible in $\Gamma .$ $\square $

\subsection{Proof of Theorem \protect\ref{sec:cce}.2}

To show the only-if part of Theorem \ref{sec:cce}.2, we need the following
lemmas.\smallskip \newline
\textbf{Lemma \ref{sec:cce}.5 (Respect of preferences }$\rightarrow $ $u$-%
\textbf{centered belief)}. Consider $M^{co}=(T_{i},b_{i})_{i\in I}$ and $%
M^{in}=(\Theta _{i},w_{i},\beta _{i})_{i\in I}$ corresponding to $M^{co}$
constructed by the two steps in Section \ref{sec:cce}.3. If $t_{i}\in T_{i}$
expresses common full belief in caution and respect of preferences, then
each $\theta _{i}\in \Theta _{i}(t_{i})$ expresses full belief in $u$%
-centered belief.\smallskip \newline
\textbf{Proof of the only-if part of Theorem \ref{sec:cce}.2}. Consider $%
M^{co}=(T_{i},b_{i})_{i\in I}$, $M^{in}=(\Theta _{i},w_{i},\beta _{i})_{i\in
I}$ corresponding to $M^{co}$, a properly rationalizable $c_{i}^{\ast }\in
C_{i}$, and $t_{i}^{\ast }\in T_{i}$ expressing common full belief in
caution and respect of preferences such that $c_{i}^{\ast }$ is rational for 
$t_{i}^{\ast }$. Let $\theta _{i}^{\ast }=\theta _{i1}(t_{i}^{\ast }).$
Since $w_{i}(\theta _{i}^{\ast })=u_{i}$ and $\beta _{i}(\theta _{i}^{\ast
}) $ has the same distribution on $j$'s choices as $b_{i}(t_{i}^{\ast })$, $%
c_{i}^{\ast }$ is rational for $\theta _{i}^{\ast }.$ Also, it follows from
Observations \ref{sec:cce}.3 and Lemmas \ref{sec:cce}.1 and \ref{sec:cce}.5
that $\theta _{i}^{\ast }$ expresses common belief in caution, $u$-centered
belief, and that a better choice is supported by utilities nearer to $u$. $%
\square $\smallskip

To show the if part, we need the following lemma.\smallskip \newline
\textbf{Lemma \ref{sec:cce}.6} \textbf{(Caution}$^{in}$\textbf{\ + }$u$%
\textbf{-centered belief }$+$\textbf{\ a better choice is supported by
utilities nearer to }$u\rightarrow $ \textbf{respect of preferences)}.
Consider $M^{in}=(\Theta _{i},w_{i},\beta _{i})_{i\in I}$ and $%
M^{co}=(T_{i},b_{i})_{i\in I}$ corresponding to $M^{in}$ constructed by the
approach in Section \ref{sec:cce}.3. If $\theta _{i}\in \Theta _{i}$
expresses common full belief in caution, $u$-centered belief, and that a
better choice is supported by utilities nearer to $u$, then $%
t_{i}(E_{i}(\theta _{i}))$ expresses common full belief in respect of
preferences.\smallskip \newline
\textbf{Proof of the if part of Theorem \ref{sec:cce}.2}. Consider $%
M^{co}=(T_{i},b_{i})_{i\in I}$ corresponding to $M^{in}$ and $c_{i}^{\ast
}\in C_{i}$ be rational for some $\theta _{i}^{\ast }$ with $w_{i}(\theta
_{i}^{\ast })=u_{i}$ which expresses common belief in caution, rationality, $%
u$-centered belief, and that a better choice is supported by utilities
nearer to $u$. Consider $t_{i}(E_{i}(\theta _{i}^{\ast })).$ Since $%
w_{i}(\theta _{i}^{\ast })=u_{i}$ and $t_{i}(E_{i}(\theta _{i}^{\ast }))$
and $\theta _{i}^{\ast }$ have the same distribution on $j$'s choices in
each level, $c_{i}^{\ast }$ is rational for $t_{i}(E_{i}(\theta _{i}^{\ast
})).$ Also, it follows from Lemmas \ref{sec:cce}.3 and \ref{sec:cce}.6 that $%
t_{i}(E_{i}(\theta _{i}^{\ast }))$ expresses common full belief in caution
and respect of preferences. Hence $c_{i}^{\ast }$ is properly rationalizable
in $\Gamma .$ $\square $

\subsection{Proof of Theorem \protect\ref{sec:cce}.3}

The only-if part of Theorem \ref{sec:cce}.3 follows directly from
Observation \ref{sec:cce}.2, Lemma \ref{sec:cce}.1, and the following lemma
whose proof can be found in Section \ref{sec:pro}.\smallskip \newline
\textbf{Lemma \ref{sec:cce}.7 (Primary belief in rationality }$\rightarrow $%
\textbf{\ Primary belief in }$u$\textbf{)}. Consider $%
M^{co}=(T_{i},b_{i})_{i\in I}$ and $M^{in}=(\Theta _{i},w_{i},\beta
_{i})_{i\in I}$ corresponding to $M^{co}$. If $t_{i}\in T_{i}$ expresses
common full belief in primary belief in rationality, then each $\theta
_{i}\in \Theta _{i}(t_{i})$ expresses common full belief in primary belief
in $u$.\smallskip

To show the if part, we need the following lemma.\smallskip \newline
\textbf{Lemma \ref{sec:cce}.8} \textbf{(Rationality\ }$+$ \textbf{primary
belief in }$u\rightarrow $\textbf{\ Primary belief in rationality)}.
Consider $M^{in}=(\Theta _{i},w_{i},\beta _{i})_{i\in I}$ and $%
M^{co}=(T_{i},b_{i})_{i\in I}$ corresponding to $M^{in}$. If $\theta _{i}\in
\Theta _{i}$ expresses common full belief in rationality and primary belief
in $u$, then $t_{i}(E_{i}(\theta _{i}))$ expresses common full belief in
primary belief in rationality.\smallskip \newline
\textbf{Proof of the if part of Theorem \ref{sec:cce}.3}. Consider $%
M^{co}=(T_{i},b_{i})_{i\in I}$ corresponding to $M^{in}$ and $c_{i}^{\ast
}\in C_{i}$ rational for some $\theta _{i}^{\ast }$ with $w_{i}(\theta
_{i}^{\ast })=u_{i}$ which expresses common full belief in caution,
rationality, and primary belief in $u$. Consider $t_{i}(E_{i}(\theta
_{i}^{\ast })).$ Since $w_{i}(\theta _{i}^{\ast })=u_{i}$ and $%
b_{i}(t_{i}(E_{i}(\theta _{i}^{\ast })))$ has the same distribution on $j$'s
choices at each level as $\beta _{i}(\theta _{i}^{\ast })$, $c_{i}^{\ast }$
is rational for $t_{i}(E_{i}(\theta _{i}^{\ast })).$ By Lemmas \ref{sec:cce}%
.3 and \ref{sec:cce}.8, $t_{i}(E_{i}(\theta _{i}^{\ast }))$ expresses common
full belief in\ caution and primary belief in rationality. Hence $%
c_{i}^{\ast }$ is permissible in $\Gamma .$ $\square $

\subsection{Proof of Theorem \protect\ref{sec:cce}.4}

\textbf{Lemma \ref{sec:cce}.9 (Assumption of rationality }$\rightarrow $%
\textbf{\ every good choice is supported\ + prior assumption of }$u,$ 
\textbf{every good choice is supported\ + prior assumption of }$u$\textbf{\
+ rationality }$\rightarrow $\textbf{\ assumption of rationality)}. Consider 
$M^{co}=(T_{i},b_{i})_{i\in I}$ and $M^{in}=(\Theta _{i},w_{i},\beta
_{i})_{i\in I}$ corresponding to $M^{co}$. If $t_{i}\in T_{i}$ expresses
common assumption of rationality, then each $\theta _{i}\in \Theta
_{i}(t_{i})$ expresses common assumption of prior $u$ and that every good
choice is supported.

On the other hand, consider $M^{in}=(\Theta _{i},w_{i},\beta _{i})_{i\in I}$
and $M^{co}=(T_{i},b_{i})_{i\in I}$ corresponding to $M^{in}$. If $\theta
_{i}\in \Theta _{i}$ expresses common full belief in rationality and\textrm{%
\ }common assumption of prior $u$ and that every good choice is supported,
then $t_{i}(E_{i}(\theta _{i}))$ expresses common assumption of
rationality.\smallskip

Here we have to combine the two ways (i.e., complete information models to
incomplete information ones and the other way around) into one lemma. The
reason is that to prove Lemma \ref{sec:cce}.9, we cannot separate them as we
did in the previous lemmas; we need to show two ways in the induction base
as well as in the induction step. The details are left to Section \ref%
{sec:pro}.\smallskip \newline
\textbf{Proof of Theorem \ref{sec:cce}.4}. \textbf{(Only-if) }Consider $%
M^{co}=(T_{i},b_{i})_{i\in I}$, $M^{in}=(\Theta _{i},w_{i},\beta _{i})_{i\in
I}$ corresponding to $M^{co},$ an iteratively admissible choice $c_{i}^{\ast
}\in C_{i}$, and $t_{i}^{\ast }\in T_{i}$ expressing common full belief in
caution and common assumption of rationality such that $c_{i}^{\ast }$ is
rational for $t_{i}^{\ast }.$ Let $\theta _{i}^{\ast }=\theta
_{i1}(t_{i}^{\ast }).$ By definition, $w_{i}(\theta _{i}^{\ast })=u_{i}$ and 
$\beta _{i}(\theta _{i}^{\ast })$ has the same distribution on $j$'s choices
at each level as $b_{i}(t_{i}^{\ast })$. Hence $c_{i}^{\ast }$ is rational
for $\theta _{i}^{\ast }.$ Also, it follows from Observation \ref{sec:cce}.2
and Lemmas \ref{sec:cce}.1 and \ref{sec:cce}.9 that $\theta _{i}^{\ast }$
expresses common full belief in caution, rationality, and common assumption
of prior $u$ and that a good choice is supported.

\textbf{(If)}. Consider $M^{co}=(T_{i},b_{i})_{i\in I}$ corresponding to $%
M^{in}$ and $c_{i}^{\ast }\in C_{i}$ rational for some $\theta _{i}^{\ast }$
with $w_{i}(\theta _{i}^{\ast })=u_{i}$ which expresses common full belief
in caution, rationality, and common assumption of prior $u$ and that a good
choice is supported. Consider $t_{i}(E_{i}(\theta _{i}^{\ast })).$ Since $%
w_{i}(\theta _{i}^{\ast })=u_{i}$ and $b_{i}(t_{i}(E_{i}(\theta _{i}^{\ast
})))$ has the same distribution on $j$'s choices at each level as $\beta
_{i}(\theta _{i}^{\ast })$, $c_{i}^{\ast }$ is rational for $%
t_{i}(E_{i}(\theta _{i}^{\ast })).$ By Lemmas \ref{sec:cce}.2 and \ref%
{sec:cce}.9, $t_{i}(E_{i}(\theta _{i}^{\ast }))$ expresses common full
belief in caution and common assumption of rationality. Therefore, $%
c_{i}^{\ast }$ is iteratively admissible. $\square $

\section{Concluding Remarks\label{sec:cr}}

\subsection{Faithful parallel to Perea and Roy (2017)'s Theorem
6.1}

All theorems here can be rephrased as faithful parallels to Perea and Roy 
(2017)'s Theorem 6.1, focusing on equivalence between belief
hierarchies in complete and incomplete information models. We adopt the
forms here because the coincidence of belief hierarchies holds by
construction, and we think it is unnecessary to mention it independently.

Also, our proofs are based on constructing a specific correspondence between
two models. It can be seen that this correspondence can be translated
directly into probabilistic models and be used to show Perea and Roy (2017)'s Theorem 6.1. Further, it can be seen that, by using our Lemma \ref%
{sec:asu}.1, belief in rationality under closest utility function in Perea
and Roy (2017) can be replaced by the weaker one (Definition \ref%
{sec:asu}.4 (4.2)) here.

\subsection{Extending to $n$-person cases}

Both Perea and Roy (2017) and this paper focus on $2$-person games. To
extend those results to $n$-person cases, the problem is how to define the
distance between utility functions and how to relate the distance with the
locations of choice-type pairs. In a $2$-person game, a type of $i$ only
needs to consider distributions on $\Delta (C_{j}\times \Theta _{j}).$ Hence
a \textquotedblleft cell\textquotedblright\ in $\beta _{i}(\theta _{i})$ is
just a pair $(c_{j},\theta _{j}),$ and its location in $\beta _{i}(\theta
_{i})\,$\ can be related directly to the distance $d(w_{j}(\theta
_{j}),u_{j}).$ In contrast, in an $n$-person case a \textquotedblleft
cell\textquotedblright\ of a lexicographic belief contains $n-1$ pairs such
as
\begin{equation*}
\left\langle (c_{1},\theta _{1}),...,(c_{i-1},\theta
_{i-1}),(c_{i+1},\theta _{i+1}),...,(c_{n},\theta _{n})\right\rangle,
\end{equation*}
and consequently there are $n-1$ distances, i.e., 
\begin{equation*}
d(w_{1}(\theta_{1}),u_{1}),...,d(w_{i-1}(\theta _{i-1}),u_{i-1}),d(w_{i+1}(\theta
_{i+1}),u_{i+1}), ... ,d(w_{n}(\theta _{n}),u_{n}).
\end{equation*}
Then the problem is
how to connect the location of this cell and those distances. We believe
that the results of Perea and Roy (2017) and this paper can be extended
to $n$-person games with a proper definition of the distances and their
relation with locations of \textquotedblleft cells\textquotedblright\ in
lexicographic beliefs. Further work is expected in that direction.

\subsection{The role of rationality}

Rationality has not been used in Theorems \ref{sec:cce}.1 and \ref{sec:cce}%
.2 even though in epistemic models with incomplete information constructed
in Section \ref{sec:cce}.3 each type has a common full belief in rationality
(Observation \ref{sec:cce}.2). On the other hand, there are also epistemic
models with types satisfying all conditions in Theorems \ref{sec:cce}.1 and %
\ref{sec:cce}.2 but not believing in rationality. Here is an
example.\smallskip \newline
\textbf{Example \ref{sec:cr}.1 (Rationality is not satisfied)}. Consider the
game $\Gamma $ in Example \ref{sec:cce}.1 and the lexicographic epistemic
model $M^{in}=(\Theta _{i},w_{i},\beta _{i})_{i\in I}$ with incomplete
information for the corresponding game form where $\Theta _{1}=\{\theta
_{11},\theta _{12},\theta _{13}\},$ $\Theta _{2}=\{\theta _{21},\theta
_{22},\theta _{23}\},$ and%
\begin{eqnarray*}
w_{1}(\theta _{11}) &=&u_{1},\text{ }\beta _{1}(\theta _{11})=((D,\theta
_{21}),(F,\theta _{22}),(E,\theta _{23})), \\
w_{1}(\theta _{12}) &=&v_{1},\text{ }\beta _{1}(\theta _{12})=((D,\theta
_{21}),(F,\theta _{22}),(E,\theta _{23})), \\
w_{1}(\theta _{13}) &=&v_{1}^{\prime },\text{ }\beta _{1}(\theta
_{13})=((D,\theta _{21}),(F,\theta _{22}),(E,\theta _{23})), \\
w_{2}(\theta _{21}) &=&v_{2},\text{ }\beta _{2}(\theta _{21})=((C,\theta
_{11}),(B,\theta _{12})),(A,\theta _{13})), \\
w_{2}(\theta _{22}) &=&v_{2}^{\prime },\text{ }\beta _{2}(\theta
_{22})=((C,\theta _{11}),(B,\theta _{12})),(A,\theta _{13})), \\
w_{2}(\theta _{23}) &=&v_{2}^{\prime \prime },\text{ }\beta _{2}(\theta
_{23})=((C,\theta _{11}),(B,\theta _{12})),(A,\theta _{13})).
\end{eqnarray*}%
where $v_{1},v_{1}^{\prime },v_{2},v_{2}^{\prime }$ are the same as in
Example \ref{sec:cce}.1 and $v_{2}^{\prime \prime }$ is illustrated in Table 11.%
\begin{table}
\label{teb:11}
\caption{An alternative utility function for player $2$}
\begin{tabular}{llll}
\hline \noalign{\smallskip}
$v_{2}^{\prime\prime}$ & $D$ & $E$ & $F$ \\ 
\noalign{\smallskip} \hline \noalign{\smallskip}
$A$ & $3$ & $2$ & $1$ \\ 
$B$ & $3$ & $2$ & $1$\\ 
$C$ & $6$ & $4$ & $5$\\
\noalign{\smallskip} \hline
\end{tabular}
\end{table}

It can be seen that $\theta _{11}$ expresses common full belief in caution, $%
u$-centered belief and that a better choice is supported by utilities nearer
to $u$ (therefore primary belief in utilities nearest to $u$ and that a best
choice is supported by utilities nearest to $u$ are also satisfied) but not
rationality since, for example, $D$ is not rational for $\theta _{21}$.
However, consider the model $M^{co}=(T_{i},b_{i})_{i\in I}$ for $\Gamma $
constructed from $M^{in}$. Indeed, since $\mathbb{E}_{1}=\{\{\theta
_{11},\theta _{12},\theta _{13}\}\}$ and $\mathbb{E}_{2}=\{\{\theta
_{21},\theta _{22},\theta _{23}\}\}$, by letting $t_{1}=t_{1}(\{\theta
_{11},\theta _{12},\theta _{13}\})$ and $t_{2}=t_{2}(\{\theta _{21},\theta
_{22},\theta _{23}\}),$ we obtain $M^{co}=(T_{i},b_{i})_{i\in I}$ for $%
\Gamma $ where $T_{1}=\{t_{1}\},$ $T_{2}=\{t_{2}\}$, and%
\begin{equation*}
b_{1}(t_{1})=((D,t_{2}),(F,t_{2}),(E,t_{2})),\text{ }%
b_{2}(t_{2})=((C,t_{1}),(B,t_{1}),(A,t_{1})).
\end{equation*}%
It can be seen that $t_{1}$ expresses caution and common full belief in
respect of preferences (therefore primary belief in rationality). Further, $%
C $ is optimal for both $\theta _{11}$ and $t_{1}$.\smallskip

On the other hand, rationality plays a critical role in the
characterizations of the second group. It seems that whether or not using
rationality in the characterization differentiates the two refinements of
permissibility within the incomplete information framework, which
corresponds to the fact that, in complete information models, there is no
general relationship between respect of preferences and assumption of
rationality (Perea, 2012). It would be interesting that any future
research could confirm this statement or show that proper rationalizability
can be characterized with rationality while iterated admissibility can be
done without it.

On the other hand, as shown in the construction in Section \ref{sec:cce}.3,
it is always possible to construct epistemic models with incomplete
information which satisfies rationality as well as all conditions in
Theorems \ref{sec:cce}.1 and \ref{sec:cce}.2. Further, prior belief in $u$
is a condition between primary belief in $u$ and $u$-centered belief. Those
seem correspond to the fact within the complete information framework that
there is always possible to construct belief hierarchy which both assumes
the opponent's rationality and respects the opponent's preferences (Perea, 2012).

\subsection{An ordinal distance on $V_{i}$}

In this note, we use the Euclidean distance to measure similarity between
utility functions. As mentioned in Section \ref{sec:asu}.1, the Euclidean
distance is cardinal. We can define an ordinal distance as follows to
replace it. Let $\beta _{i}$ be a lexicographic belief on $\Delta
(S_{j}\times \Theta _{j}).$ For each $v_{i},u_{i}\in V_{i},$ define $%
d^{\beta _{i}}(v_{i},u_{i})=|\{\{s_{i},s_{i}^{\prime }\}:s_{i},s_{i}^{\prime
}\in S_{i}$ and the preference between $s_{i}$ and $s_{i}^{\prime }$ under $%
\beta _{i}$ at $v_{i}$ are different from that at $u_{i}\}|.$ It can be seen
that $d^{\beta _{i}}$ is a variation of Hamming distance (Hamming, 1950). It measures similarity between preferences under $\beta _{i}$ represented
by $v_{i}$ and that by $u_{i},$ i.e., it measures the ordinal difference
between $v_{i}$ and $u_{i}.$ This does not belong to the group of distances
characterized in Section 3.3 of Perea and Roy (2017) since there is no
norm on $V_{i}$ to support $d^{\beta _{i}}.$ Lemma \ref{sec:asu}.1 still
holds under $d^{\beta _{i}}$ since even if we replace $d$ by $d^{\beta _{i}}$
in Lemma \ref{sec:asu}.1 (c), the constructed utility function sequence in
the proof still satisfies it. Hence $d$ in Definition \ref{sec:asu}.4 can be
replaced by $d^{\beta _{i}}$ with appropriate $\beta _{i}$ and the
characterization results still hold. Also, by replacing rationality under
closest utility function by our Definition \ref{sec:asu}.4 (4.2), Perea and
Roy (2017)'s Theorem 6.1 still holds under $d^{\beta _{i}}.$

\subsection{Weakening caution in Theorems \protect\ref{sec:cce}.3 and 
\protect\ref{sec:cce}.4}

Caution in Theorems \ref{sec:cce}.3 and \ref{sec:cce}.4 can be replaced by a
weaker concept called \textquotedblleft weak caution\textquotedblright\
which is defined as follows.\smallskip \newline
\textbf{Definition \ref{sec:cr}.1 (Weak caution)}. Consider a game form $%
G=(C_{i})_{i\in I}$ and a lexicographic epistemic model $M^{in}=(\Theta
_{i},w_{i},\beta _{i})_{i\in I}$ for $G$ with incomplete information. $%
\theta _{i}\in \Theta _{i}$ is \emph{weakly cautious} iff for each $c_{j}\in
C_{j},$ there is some $\theta _{j}\in \Theta _{j}$ such that $\theta _{i}$
deems $(c_{j},\theta _{j})$ possible.\smallskip

Definition \ref{sec:cr}.1 is weaker than Definition \ref{sec:asu}.2 since it
only requires that each choice should appear in the belief of $\theta _{i}$
but does not require that it should be paired with each belief of $j$ deemed
possible by $\theta _{i}$. Nevertheless, we will show in Lemma \ref{sec:cr}%
.1 that in with other conditions in this characterization it leads to
caution.\smallskip \newline
\textbf{Proposition \ref{sec:cr}.1 (An alternative characterization of
permissibility)}. Consider a finite 2-person static game $\Gamma
=(C_{i},u_{i})_{i\in I},$ the corresponding game form $G=(C_{i})_{i\in I}$,
and a finite lexicographic epistemic model $M^{co}=(T_{i},b_{i})_{i\in I}$
for $\Gamma .$

Then, $c_{i}^{\ast }\in C_{i}$ is permissible in $M^{co}$ if and only if
there is some finite lexicographic epistemic model $M^{in}=(\Theta
_{i},w_{i},\beta _{i})_{i\in I}$ with incomplete information for $G$ and
some $\theta _{i}^{\ast }\in \Theta _{i}$ with $w_{i}(\theta _{i}^{\ast
})=u_{i}$ such that\smallskip \newline
(a) $c_{i}^{\ast }$ is rational for $\theta _{i}^{\ast },$ and,\smallskip 
\newline
(b) $\theta _{i}^{\ast }$ expresses common full belief in weak caution,
rationality, and primary belief in $u$.\smallskip

The only-if part holds since weak caution is weaker than caution. The if
part needs first to show that weak caution is enough for the
characterization. Here, we show that the corresponding concept in complete
information model can replace caution and characterize permissibility. Then
we can use the mapping between complete and incomplete information models
constructed in Section \ref{sec:cce}.3. Let $M^{co}=(T_{i},b_{i})_{i\in I}$
be a lexicographic model for $\Gamma =(C_{i},u_{i})_{i\in I}$ with complete
information. $t_{i}\in T_{i}$ is \emph{weakly cautious} iff for each $%
c_{j}\in C_{j},$ there is some $t_{j}\in T_{j}$ such that $t_{i}$ deems $%
(c_{j},t_{j})$ possible. We have the following lemma.\smallskip \newline
\textbf{Lemma \ref{sec:cr}.1 (Characterizing permissibility by weak
caution). }Consider a lexicographic epistemic model $M^{co}=(T_{i},b_{i})_{i%
\in I}$ for a game $\Gamma =(C_{i},u_{i})_{i\in I}$. $c_{i}^{\ast }\in C_{i}$
is permissible if and only if it is rational to some $t_{i}^{\ast }\in T_{i}$
which expresses common full belief in weak caution and primary belief in
rationality.\smallskip

Also, we need the following lemmas.\smallskip \newline
\textbf{Lemma \textbf{\ref{sec:cr}}.2} \textbf{(Weak caution}$%
^{in}\rightarrow $\textbf{\ weak caution}$^{co}$\textbf{)}. Let $%
M^{in}=(\Theta _{i},w_{i},\beta _{i})_{i\in I}$ and $M^{co}=(T_{i},b_{i})_{i%
\in I}$ be constructed from $M^{in}$ by the above approach. If $\theta
_{i}\in \Theta _{i}$ expresses common full belief in weak caution, so does $%
t_{i}(E_{i}(\theta _{i})).$\smallskip

We omit the proof of Lemma \ref{sec:cr}.2 since it can be shown in a similar
way as Lemma \ref{sec:cce}.3. It can be seen that Proposition \ref{sec:cr}.1
follows directly from Lemmas \ref{sec:cr}.1 and \ref{sec:cr}.1 and Theorem %
\ref{sec:cce}.3. Similarly, it can be seen that caution in Theorem \ref%
{sec:cce}.4 can be replaced by weak caution.

However, it should be noted that caution cannot be weakened in Theorems \ref%
{sec:cce}.1 and \ref{sec:cce}.2. For Theorem \ref{sec:cce}.1, caution plays
an important role in the proof of the if part; without it, primary belief in
utilities nearest to $u$ and that a best choice is supported by utilities
nearest to $u$ cannot imply primary belief in rationality. For Theorem \ref%
{sec:cce}.2, the interpolation method used in the proof of Lemma \ref{sec:cr}%
.1 does not work in general since different types may have different orders
there.

\section{Proofs}
\label{sec:pro}

\textbf{Proof of Lemma \ref{sec:asu}.1}. We construct a sequence satisfying
(a)-(c) by induction. First, let $v_{i1}=u_{i}.$ Suppose that for some $\ell
\in \{1,...,L-1\}$ we have defined $v_{i1},...,v_{i\ell }$ satisfying
(a)-(c). Now we show how to define \thinspace $v_{i,\ell +1}.$ It can be
seen that there exists $M_{\ell +1}>0$ such that $v_{i\ell }(c_{i,\ell
+1},\beta _{i1})+M_{\ell +1}>v_{i\ell }(c_{i\ell },\beta _{i1})$ for all $%
c_{i\ell }\in C_{i\ell }$\ and $c_{i,\ell +1}\in C_{i,\ell +1}.$ We define $%
v_{i,\ell +1}$ as follows: for each $(c_{i},c_{j})\in C,$%
\begin{equation*}
v_{i,\ell +1}(c_{i},c_{j})=\left\{ 
\begin{array}{c}
v_{i\ell }(c_{i},c_{j})+M_{\ell +1}\text{ \ if }c_{i}\in C_{i,\ell +1}\text{
and }c_{j}\in \text{supp}\beta _{i1} \\ 
v_{i\ell }(c_{i},c_{j})\text{ \ \ \ \ \ \ \ \ \ \ \ \ \ \ \ \ \ \ \ \ \ \ \
\ \ \ \ \ \ \ \ \ \ \ \ \ \ \ \ otherwise}%
\end{array}%
\right.
\end{equation*}%
It can be seen that each $c_{i,\ell +1}\in C_{i,\ell +1}$ is rational at $%
v_{i,\ell +1}$ under $\beta _{i}.$ Also, since $d(v_{i,\ell +1},v_{i\ell
})=(M_{\ell +1}^{2}\times |C_{i,\ell +1}|\times |$supp$\beta _{i1}|)^{1/2}>0$%
, $d(v_{i,\ell +1},u_{i})=d(v_{i,\ell +1},v_{i\ell
})+d(v_{in},u_{i})>d(v_{in},u_{i}).$ By induction, we can obtain a sequence $%
v_{i1},...,v_{iL}\in V_{i}$ satisfying (a)-(c). $\square $\smallskip

It should be noted that, given $u_{i}$ and $\beta _{i},$ the sequence $%
v_{i1},...,v_{iL}$ satisfying (a)-(c) is not unique. The basic idea behind
this inductive construction is depicted as follows. Suppose that $%
u_{i}(c_{i1},\beta _{i})>u_{i}(c_{i2},\beta _{i})>...>u_{i}(c_{iN},\beta
_{i}),$ that is, $\Pi _{i}(\beta
_{i})=(\{c_{i1}\},\{c_{i2}\},...,\{c_{iN}\}) $, then%
\begin{equation*}
\left( c_{i1},c_{i2},c_{i3},...,c_{iN}\right) \text{ \ }\underrightarrow{%
v_{i2}}\text{ \ }\left( c_{i2},c_{i1},c_{i3},...,c_{iN},\right) \text{ \ }%
\underrightarrow{v_{i3}}\text{ \ }\left(
c_{i3},c_{i2},c_{i1}...,c_{iN}\right) \text{ \ ... \ }\underrightarrow{v_{iN}%
}\text{ \ }\left( c_{iN},c_{i,N-1},...,c_{i1}\right)
\end{equation*}%
Informally speaking, we take equivalent classes of choices one by one to the
foremost location of the sequence according to the order of preference in $%
u_{i}$ under $\beta _{i}$. The following example shows how this construction
works.\smallskip \newline
\textbf{Example \ref{sec:pro}.1.} Consider $u_{1}$ in Example \ref{sec:mac}%
.1. Under the lexicographic belief $\beta _{1}=(D,E,F)$, $A$ is preferred to 
$B$ and $B$ is preferred to $C$ in $u_{1},$ that is, $\Pi _{1}(\beta
_{1})=(\{A\},\{B\},\{C\}).$ We can define $v_{11},v_{12}$, and $v_{13}$ as in Table 12.%
\begin{table}
\label{tab:12}
\caption{Alternative utility functions for players $1$ corresponding to $\Pi _{1}(\beta
_{1})$}
\begin{tabular}{llll}
\hline \noalign{\smallskip}
$v_{11}=u_{1}$ & $D$ & $E$ & $F$ \\ 
\noalign{\smallskip} \hline \noalign{\smallskip}
$A$ & $1$ & $1$ & $1$ \\ 
$B$ & $1$ & $1$ & $0$\\ 
$C$ & $1$ & $0$ & $1$\\
\noalign{\smallskip} \hline
\end{tabular}\quad $\longrightarrow$ \quad
\begin{tabular}{llll}
\hline \noalign{\smallskip}
$v_{12}$ & $D$ & $E$ & $F$ \\ 
\noalign{\smallskip} \hline \noalign{\smallskip}
$A$ & $1$ & $1$ & $1$ \\ 
$B$ & $2$ & $1$ & $0$\\ 
$C$ & $1$ & $0$ & $1$\\
\noalign{\smallskip} \hline
\end{tabular}\quad $\longrightarrow$ \quad
\begin{tabular}{llll}
\hline \noalign{\smallskip}
$v_{13}$ & $D$ & $E$ & $F$ \\ 
\noalign{\smallskip} \hline \noalign{\smallskip}
$A$ & $1$ & $1$ & $1$ \\ 
$B$ & $2$ & $1$ & $0$\\ 
$C$ & $3$ & $0$ & $1$\\
\noalign{\smallskip} \hline
\end{tabular}\quad 
\end{table}

At $v_{11},$ the order of preferences is $(A,B,C)$ under $\beta _{1},$ at $%
v_{12}$ it is $(B,A,C),$ and at $v_{13}$ it is $(C,B,A).$\smallskip \newline
\textbf{Proof of Lemma \ref{sec:cce}.1}. We show this statement by
induction. First we show that if $t_{i}$ is cautious, then each $\theta
_{i}\in \Theta _{i}(t_{i})$ is also cautious. Let $c_{j}\in C_{j}$ and $%
\theta _{j}\in \Theta _{j}(\theta _{i}).$ By construction, it can be seen
that the type $t_{j}\in T_{j}$ satisfying the condition that $\theta _{j}\in
\Theta _{j}(t_{j})$ is in $T_{j}(t_{i}).$ Since $t_{i}$ is cautious, $t_{i}$
deems $(c_{j},t_{j})$ possible. Consider the pair $(c_{j},\theta
_{j}^{\prime })$ in $\beta _{i}(\theta _{i})$ corresponding to $%
(c_{j},t_{j}).$ Since both $\theta _{j}$ and $\theta _{j}^{\prime }$ are in $%
\Theta _{j}(t_{j}),$ it follows from Observation \ref{sec:cce}.1 that $\beta
_{j}(\theta _{j})=\beta _{j}(\theta _{j}^{\prime }).$ Hence $(c_{j},\theta
_{j}^{w_{j}(\theta _{j}^{\prime })})$ is deemed possible by $\theta _{i}.$
Here we have shown that $\theta _{i}$ is cautious.

Suppose we have shown that, for each $i\in I,$ if $t_{i}$ expresses $n$-fold
full belief in caution then so does each $\theta _{i}\in \Theta _{i}(t_{i})$%
. Now suppose that $t_{i}$ expresses $(n+1)$-fold full belief in caution,
i.e., each $t_{j}\in T_{j}(t_{i})$ expresses $n$-fold full belief in
caution. By construction, for each $\theta _{i}\in \Theta _{i}(t_{i})$ and
each $\theta _{j}\in \Theta _{j}(\theta _{i})$ there is some $t_{j}\in
T_{j}(t_{i})$ such that $\theta _{j}\in \Theta _{j}(t_{i})$, and, by
inductive assumption, each $\theta _{j}\in \Theta _{j}(\theta _{i})$
expresses $n$-fold full belief in caution. Therefore, each $\theta _{i}\in
\Theta _{i}(t_{i})$ expresses $(n+1)$-fold full belief in caution. $\square $%
\smallskip \newline
\textbf{Proof of Lemma \ref{sec:cce}.2}. We show this statement by
induction. First we show that if $t_{i}$ primarily believes in $j$'s
rationality, then each $\theta _{i}\in \Theta _{i}(t_{i})$ primarily
believes in utilities nearest to $u$. Let $(c_{j},\theta _{j})$ be a pair
deemed possible in the level-1 belief of $\theta _{i}.$ Consider its
correspondence $(c_{j},t_{j})$ in level-1 belief of $t_{i}.$ Since $t_{i}$
primarily believes in $j$'s rationality, $c_{j}$ is rational for $t_{j}.$ It
follows that $c_{j}\in C_{j1}\in \Pi _{j}(t_{j}).$ By Lemma \ref{sec:mac}.1
and construction, it follows that $w_{j}(\theta _{j})=u_{j}.$ Since $u_{j}$
is the nearest function to itself among all utility functions in $V_{j},$ we
have shown that $\theta _{i}$ primarily believes in utilities nearest to $u.$

Suppose we have shown that, for each $i\in I,$ if $t_{i}$ expresses $n$-fold
full belief in primary belief in rationality then each $\theta _{i}\in
\Theta _{i}(t_{i})$ expresses $n$-fold full belief in primary belief in
utilities nearest to $u$. Now suppose that $t_{i}$ expresses $(n+1)$-fold
full belief in primary belief in rationality, i.e., each $t_{j}\in
T_{j}(t_{i})$ expresses $n$-fold full belief in primary belief in
rationality. Since, by construction, for each $\theta _{i}\in \Theta
_{i}(t_{i})$ and each $\theta _{j}\in \Theta _{j}(\theta _{i})$ there is
some $t_{j}\in T_{j}(t_{i})$ such that $\theta _{j}\in \Theta _{j}(t_{j})$,
it follows that, by inductive assumption, each $\theta _{j}\in \Theta
_{j}(\theta _{i})$ expresses $n$-fold full belief in primary belief in
utilities nearest to $u$. Therefore, each $\theta _{i}\in \Theta _{i}(t_{i})$
expresses $(n+1)$-fold full belief in primary belief in utilities nearest to 
$u$. $\square $\smallskip \newline
\textbf{Proof of Lemma \textbf{\ref{sec:cce}}.3}. We show this statement by
induction. First we show that if $\theta _{i}$ is cautious, then $%
t_{i}(E_{i}(\theta _{i}))$ is also cautious. Let $c_{j}\in C_{j}$ and $%
t_{j}\in T_{j}(t_{i}(E_{i}(\theta _{i}))).$ By construction, $%
t_{j}=t_{j}(E_{j})$ for some $E_{j}\in \mathbb{E}_{j},$ and there is some $%
\theta _{j}\in E_{j}$ which is deemed possible by $\theta _{i}.$ Since $%
\theta _{i}$ is cautious, there is some $\theta _{j}^{\prime }$ with $\beta
_{j}(\theta _{j}^{\prime })=\beta _{j}(\theta _{j}),$ i.e., $\theta
_{j}^{\prime }\in E_{j},$ such that $(c_{j},\theta _{j}^{\prime })$ is
deemed possible by $\theta _{i}.$ By construction, $(c_{j},t_{j})$ is deemed
possible by $t_{i}(E_{i}(\theta _{i})).$

Suppose we have shown that, for each $i\in I,$ if $\theta _{i}$ expresses $n$%
-fold full belief in caution then so does $t_{i}(E_{i}(\theta _{i}))$. Now
suppose that $\theta _{i}$ expresses $(n+1)$-fold full belief in caution,
i.e., each $\theta _{j}\in \Theta _{j}(\theta _{i})$ expresses $n$-fold full
belief in caution. Since, by construction, for each $t_{j}\in
T_{j}(t_{i}(E_{i}(\theta _{i})))$, there is some $\theta _{j}\in \Theta
_{j}(\theta _{i})$ such that $t_{j}=t_{j}(E_{j}(\theta _{j})),$ by inductive
assumption $t_{j}$ expresses $n$-fold full belief in caution. Therefore, $%
t_{i}(E_{i}(\theta _{i}))$ expresses $(n+1)$-fold full belief in caution. $%
\square $\smallskip \newline
\textbf{Proof of Lemma \ref{sec:cce}.4}. We show this statement by
induction. First we show that if $\theta _{i}$ is cautious, primarily
believes in utilities nearest to $u,$ and believes in that a best choice is
supported by utilities nearest to $u$, then $t_{i}(E_{i}(\theta _{i}))$
primarily believes in $j$'s rationality. Let $(c_{j},t_{j})$ be a
choice-type pair which is deemed possible in $t_{i}(E_{i}(\theta _{i}))$'s
level-1 belief. By construction $t_{j}=t_{j}(E_{j})$ for some $E_{j}\in 
\mathbb{E}_{j},$ and for some $\theta _{j}\in E_{j},$ $(c_{j},\theta _{j})$
is deemed possible in $\theta _{i}$'s level-1 belief. Since $\theta _{i}$
primarily believes in utilities nearest to $u,$ it follows that 
\begin{equation}
d(w_{j}(\theta _{j}),u_{j})\leq d(w_{j}(\theta _{j}^{\prime }),u_{j})\text{
for all }\theta _{j}^{\prime }\in E_{j}.  \label{op1}
\end{equation}%
Suppose that $c_{j}$ is not optimal for $t_{j}.$ Let $c_{j}^{\prime }$ be a
choice optimal to $t_{j}.$ Since $\theta _{i}$ is cautious, there is some $%
\theta _{j}^{v_{j}}\in E_{j}$ such that $(c_{j},\theta _{j}^{v_{j}})$ is
deemed possible by $\theta _{i}.$ Then since $\theta _{i}$ believes in that
a best choice is supported by utilities nearest to $u,$ it follows that $%
d(v_{j},u_{j})=d(w_{j}(\theta _{j}^{v_{j}}),u_{j})<d(w_{j}(\theta
_{j}),u_{j}),$ which is contradictory to (\ref{op1}). Therefore $c_{j}$ is
optimal for $t_{j}$. Here we have shown that $t_{i}(E_{i}(\theta _{i}))$
primarily believes in $j$'s rationality.

Suppose we have shown that, for each $i\in I,$ if $\theta _{i}$ expresses $n$%
-fold full belief in caution, primary belief in utilities nearest to $u$,
and that a best choice is supported by utilities nearest to $u$, then $%
t_{i}(E_{i}(\theta _{i}))$ expresses $n$-fold belief in primary belief in
rationality. Now suppose that $\theta _{i}$ expresses $(n+1)$-fold full
belief in caution, primary belief in utilities nearest to $u$, and that a
best choice is supported by utilities nearest to $u$, i.e., each $\theta
_{j}\in \Theta _{j}(\theta _{i})$ expresses $n$-fold full belief in caution,
primary belief in utilities nearest to $u$, and that a best choice is
supported by utilities nearest to $u$. Since, by construction, for each $%
t_{j}\in T_{j}(t_{i}(E_{i}(\theta _{i})))$, there is some $\theta _{j}\in
\Theta _{j}(\theta _{i})$ such that $t_{j}=t_{j}(E_{j}(\theta _{j})),$ by
inductive assumption $t_{j}$ expresses $n$-fold full belief in primary
belief in rationality. Therefore, $t_{i}(E_{i}(\theta _{i}))$ expresses $%
(n+1)$-fold full belief in primary belief in rationality. $\square $%
\smallskip \newline
\textbf{Proof of Lemma \ref{sec:cce}.5}. We show this statement by
induction. First we show that if $t_{i}$ is caution and respects $j$'s
preferences, then each $\theta _{i}\in \Theta _{i}(t_{i})$ expresses $u$%
-centered belief. It can be seen that if $t_{i}$ is cautious and respects $j$%
's preferences, then we can combine all types deemed possible by $t_{i}$
with the same belief into one type without hurting the caution and respect
of $j$'s preference, and every choice optimal for $t_{i}$ is still optimal
for this new type and vice versa. Therefore, without loss of generality we
can assume that for each $t_{j},t_{j}^{\prime }\in T_{j},$ $b_{j}(t_{j})\neq
b_{j}(t_{j}^{\prime }).$ Let $c_{j},c_{j}^{\prime }\in C_{j},$ $\theta
_{j}\in \Theta _{j}$, and $v_{j},v_{j}^{\prime }\in V_{j}$ such that $%
(c_{j},\theta _{j}^{v_{j}})$ and $(c_{j}^{\prime },\theta
_{j}^{v_{j}^{\prime }})$ are deemed possible by $\theta _{i}$ with $%
d(v_{j},u_{j})<d(v_{j}^{\prime },u_{j}).$ Since each type in $T_{i}$ has a
distinct lexicographic belief, it follows that $\theta _{j}^{v_{j}},\theta
_{j}^{v_{j}^{\prime }}\in \Theta _{j}(t_{j})$ for some $t_{j}\in T_{j}.$ By
construction it follows that (1) $t_{i}$ deems both $(c_{j},t_{j})$ and $%
(c_{j}^{\prime },t_{j})$ possible, and (2) $u_{j}(c_{j},t_{i})>u_{j}(c_{j}^{%
\prime },t_{i}).$ Since $t_{i}$ respects $j$'s preferences, $t_{i}$ deems $%
(c_{j},t_{j})$ infinitely more likely than $(c_{j}^{\prime },t_{j}),$ which
corresponds to that $\theta _{i}$ deems $(c_{j},\theta _{j}^{v_{j}})$
infinitely more likely than $(c_{j}^{\prime },\theta _{j}^{v_{j}^{\prime
}}). $ Here we have shown that $\theta _{i}$ expresses $u$-centered belief.

Suppose we have shown that, for each $i\in I,$ if $t_{i}$ expresses $n$-fold
full belief in respect of preferences then each $\theta _{i}\in \Theta
_{i}(t_{i})$ expresses $n$-fold full belief in $u$-centered belief. Now
suppose that $t_{i}$ expresses $(n+1)$-fold full belief in respect of
preferences, i.e., each $t_{j}\in T_{j}(t_{i})$ expresses $n$-fold full
belief respect of preferences. Since, by construction, for each $\theta
_{i}\in \Theta _{i}(t_{i})$ and each $\theta _{j}\in \Theta _{j}(\theta
_{i}) $ there is some $t_{j}\in T_{j}(t_{i})$ such that $\theta _{j}\in
\Theta _{j}(t_{j})$, by inductive assumption it follows that each $\theta
_{j}\in \Theta _{j}(\theta _{i})$ expresses $n$-fold full belief in $u$%
-centered belief. Therefore, each $\theta _{i}\in \Theta _{i}(t_{i})$
expresses $(n+1)$-fold full belief in $u$-centered belief. $\square $%
\smallskip \newline
\textbf{Proof of Lemma \ref{sec:cce}.6}. We show this statement by
induction. First we show that if $\theta _{i}$ is cautious, has a $u$%
-centered belief, and believes that a better choice is supported by
utilities nearer to $u$, then $t_{i}(E_{i}(\theta _{i}))$ respects $j$'s
preferences. First, since $\theta _{i}$ is cautious, By Lemma \ref{sec:cce}%
.3, $t_{i}(E_{i}(\theta _{i}))$ is also cautious. Let $c_{j},c_{j}^{\prime
}\in C_{j}$ and $t_{j}\in T_{j}(t_{i}(E_{i}(\theta _{i})))$ with $t_{j}$
prefers $c_{j}$ to $c_{j}^{\prime }.$ By construction $t_{j}=t_{j}(E_{j})$
for some $E_{j}\in \mathbb{E}_{j},$ and, since $\theta _{i}$ is cautious,
there are $\theta _{j},\theta _{j}^{\prime }\in E_{j}$ such that $\theta
_{i} $ deems $(c_{j},\theta _{j})$ and $(c_{j}^{\prime },\theta _{j}^{\prime
})$ possible. Since $\beta _{j}(\theta _{j})=\beta _{j}(\theta _{j}^{\prime
})$ and $\theta _{j}$ has the same probability distribution over $C_{i}$ at
each level as $t_{j},$ it follows that $u_{j}(c_{j},\theta
_{j})>u_{j}(c_{j}^{\prime },\theta _{j}).$ Since $\theta _{i}$ believes that
a better choice is supported by utilities nearer to $u,$ it follows that $%
d(w_{j}(\theta _{j}),u_{j})<d(w_{j}(\theta _{j}^{\prime }),u_{j}).$ Since $%
\theta _{i}$ has a $u$-centered belief, it follows that $\theta _{i}$ deems $%
(c_{j},\theta _{j})$ infinitely more likely than $(c_{j}^{\prime },\theta
_{j}^{\prime }),$ which implies that $t_{i}(E_{i}(\theta _{i}))$ deems $%
(c_{j},t_{j})$ infinitely more likely than $(c_{j}^{\prime },t_{j}).$
Therefore, $t_{i}(E_{i}(\theta _{i}))$ respects $j$'s preferences.

Suppose we have shown that, for each $i\in I,$ if $\theta _{i}$ expresses $n$%
-fold full belief in caution, $u$-centered belief, and that a better choice
is supported by utilities nearer to $u,$ then $t_{i}(E_{i}(\theta _{i}))$
expresses $n$-fold full belief in respect of preferences. Now suppose that $%
\theta _{i}$ expresses $(n+1)$-fold full belief in caution, $u$-centered
belief, and that a better choice is supported by utilities nearer to $u$,
i.e., each $\theta _{j}\in \Theta _{j}(\theta _{i})$ expresses $n$-fold full
belief in caution, $u$-centered belief, and that a better choice is
supported by utilities nearer to $u$. Since, by construction, for each $%
t_{j}\in T_{j}(t_{i}(E_{i}(\theta _{i})))$, there is some $\theta _{j}\in
\Theta _{j}(\theta _{i})$ such that $t_{j}=t_{j}(E_{j}(\theta _{j})),$ by
inductive assumption $t_{j}$ expresses $n$-fold full belief in respect of
preferences. Therefore, $t_{i}(E_{i}(\theta _{i}))$ expresses $(n+1)$-fold
full belief in respect of preferences. $\square $\smallskip \newline
\textbf{Proof of Lemma \ref{sec:cce}.7}. We show this statement by
induction. First we show that if $t_{i}$ primarily believes in $j$'s
rationality, then each $\theta _{i}\in \Theta _{i}(t_{i})$ primarily
believes in $u$. Let $(c_{j},\theta _{j})$ be a pair deemed possible in the
level-1 belief of $\theta _{i}.$ Consider its corresponding $(c_{j},t_{j})$
in level-1 belief of $t_{i}.$ Since $t_{i}$ primarily believes in $j$'s
rationality, $c_{j}$ is rational for $t_{j}.$ It follows that $c_{j}\in
C_{j1}\in \Pi _{j}(t_{j}).$ By construction, it follows that $w_{j}(\theta
_{j})=u_{j}.$ Here we have shown that $\theta _{i}$ primarily believes in $%
u. $

Suppose we have shown that, for each $i\in I,$ if $t_{i}$ expresses $n$-fold
full belief in primary belief in rationality then each $\theta _{i}\in
\Theta _{i}(t_{i})$ expresses $n$-fold full belief in primary belief in $u$.
Now suppose that $t_{i}$ expresses $(n+1)$-fold full belief in primary
belief in rationality, i.e., each $t_{j}\in T_{j}(t_{i})$ expresses $n$-fold
full belief in primary belief in rationality. Since, by construction, for
each $\theta _{i}\in \Theta _{i}(t_{i})$ and each $\theta _{j}\in \Theta
_{j}(\theta _{i})$ there is some $t_{j}\in T_{j}(t_{i})$ such that $\theta
_{j}\in \Theta _{j}(t_{j})$, it follows that, by inductive assumption, each $%
\theta _{j}\in \Theta _{j}(\theta _{i})$ expresses $n$-fold full belief in
primary belief in rationality. Therefore, each $\theta _{i}\in \Theta
_{i}(t_{i})$ expresses $(n+1)$-fold full belief in primary belief in $u$. $%
\square $\smallskip \newline
\textbf{Proof of Lemma \ref{sec:cce}.8}. We show this statement by
induction. First we show that if $\theta _{i}$ believes in $j$'s rationality
and primarily believes in $u$, then $t_{i}(E_{i}(\theta _{i}))$ primarily
believes in $j$'s rationality. Let $(c_{j},t_{j})$ be a choice-type pair
which is deemed possible in $t_{i}(E_{i}(\theta _{i}))$'s level-1 belief. By
construction $t_{j}=t_{j}(E_{j})$ for some $E_{j}\in \mathbb{E}_{j},$ and
for some $\theta _{j}\in E_{j},$ $(c_{j},\theta _{j})$ is deemed possible in 
$\theta _{i}$'s level-1 belief. Since $\theta _{i}$ primarily believes in $%
u, $ it follows that $w_{j}(\theta _{j})=u_{j}.$ Also, since $\theta _{i}$
believes $j$'s rationality, it follows that $c_{j}$ is rational at $u_{j}$
under $\beta _{j}(\theta _{j})$, i.e., $b_{i}(t_{j}).$ Therefore $c_{j}$ is
rational for $t_{j}$. Here we have shown that $t_{i}(E_{i}(\theta _{i}))$
primarily believes in $j$'s rationality.

Suppose we have shown that, for each $i\in I,$ if $\theta _{i}$ expresses $n$%
-fold full belief in rationality and primary belief in $u$, then $%
t_{i}(E_{i}(\theta _{i}))$ expresses $n$-fold belief in primary belief in
rationality. Now suppose that $\theta _{i}$ expresses $(n+1)$-fold full
belief in rationality and primary belief in $u$, i.e., each $\theta _{j}\in
\Theta _{j}(\theta _{i})$ expresses $n$-fold full belief in rationality and
primary belief in $u$. Since, by construction, for each $t_{j}\in
T_{j}(t_{i}(E_{i}(\theta _{i})))$, there is some $\theta _{j}\in \Theta
_{j}(\theta _{i})$ such that $t_{j}=t_{j}(E_{j}(\theta _{j})),$ by inductive
assumption $t_{j}$ expresses $n$-fold full belief in primary belief in
rationality. Therefore, $t_{i}(E_{i}(\theta _{i}))$ expresses $(n+1)$-fold
full belief in primary belief in rationality. $\square $\smallskip \newline
\textbf{Proof of Lemma \ref{sec:cce}.9}. We show this statement by
induction. Let $\theta _{i}\in \Theta _{i}(t_{i}).$ First we show that if $%
t_{i}$ assumes in $j$'s rationality, $\theta _{i}$ prior assumes $u$ and
assumes that every good choice is supported. Let $c_{j}\in C_{j}$ be optimal
for some cautious type of $j$ whose assigned utility function is $u_{j}$
within an epistemic model with incomplete information. It is easy to see
that $c_{j}$ is optimal for its corresponding type, which is also cautious
by Lemma \ref{sec:cce}.2, in any complete information model constructed from
the one with incomplete information. Since $t_{i}$ assumes $j$'s
rationality, $t_{i}$ deems possible a cautious type $t_{j}$ for which $c_{j}$
is optimal. By construction, some $\theta _{j}\in \Theta _{j}(t_{j})$ is
deemed possible by $\theta _{i}.$ Since $t_{i}$ is cautious, $(c_{j},t_{j})$
is deemed possible by $t_{i},$ and, by construction $(c_{j},\theta
_{j1}(t_{j}))$ is deemed possible by $\theta _{i}$. Since $w_{j}(\theta
_{j1}(t_{j}))=u_{j}$ and $c_{j}$ is optimal for $\theta _{j1}(t_{j}),$ it
follows that $\theta _{i}$ assumes that every good choice is supported.

Let $(c_{j},\theta _{j})$ with $\theta _{j}$ cautious deemed possible by $%
\theta _{i}$ satisfying $w_{j}(\theta _{j})=u_{j}$ and $(c_{j}^{\prime
},\theta _{j}^{\prime })$ a pair which does not satisfy that condition. Let $%
(c_{j},t_{j})$ and $(c_{j}^{\prime },t_{j}^{\prime })$ be the pairs
occurring in the belief of $t_{i}$ corresponding to $(c_{j},\theta _{j})$
and $(c_{j}^{\prime },\theta _{j}^{\prime }).$ Since $c_{j}$ is rational to $%
\theta _{j}$ and $w_{j}(\theta _{j})=u_{j},$ it follows that $c_{j}$ is
optimal for $t_{j}.$ On the other hand, $c_{j}^{\prime }$ is not optimal for 
$t_{j}^{\prime }.$ Since $t_{i}$ assumes $j$'s rationality, $t_{i}$ deems $%
(c_{j},t_{j})$ infinitely more likely than $(c_{j}^{\prime },t_{j}^{\prime
}).$ By construction, $\theta _{i}$ deems $(c_{j},\theta _{j})$ infinitely
more likely than $(c_{j}^{\prime },\theta _{j}^{\prime }).$ Here we have
shown that $\theta _{i}$ prior assumes $u.$\smallskip

Now we show the other direction: suppose that if $\theta _{i}\in \Theta _{i}$
prior assumes $u$ and assumes that every good choice is supported, we prove
that $t_{i}(E_{i}(\theta _{i}))$ assumes $j$'s rationality. Suppose that $%
c_{j}$ is optimal for some cautious type within some epistemic model with
complete information. It can be seen by construction that $c_{j}$ is optimal
for some cautious type with $u_{i}$ as its assigned utility function within
some epistemic model with incomplete information which corresponds to that
complete information model. Since $\theta _{i}$ believes in that every good
choice is supported, $\theta _{i}$ deems possible a cautious type $\theta
_{j}$ such that $w_{j}(\theta _{j})=u_{j}$ and $c_{j}$ is optimal for $%
\theta _{j}.$ By construction it follows that $t_{i}(E_{i}(\theta _{i}))$
deems $t_{j}(E_{j}(\theta _{j}))$ possible for which $c_{j}$ is optimal.

Let $(c_{j},t_{j})$ with $t_{j}$ cautious be a pair which is deemed possible
by $t_{i}(E_{i}(\theta _{i}))$ satisfying that $c_{j}$ is optimal for $%
t_{j}, $ and $(c_{j}^{\prime },t_{j}^{\prime })$ be a pair deemed possible
by $t_{i}(E_{i}(\theta _{i}))$ which does not satisfy that condition. Since $%
\theta _{i}$ assumes that every good choice is supported, there is some $%
\theta _{j}\in \Theta _{j}(\theta _{i})$ corresponding to $t_{j}$ with $%
w_{j}(\theta _{j})=u_{j}$ such that $(c_{j},\theta _{j})$ is deemed possible
by $\theta _{i}.$ On the other hand, since $\theta _{i}$ believes in
rationality, for each $\theta _{j}^{\prime }$ such that $(c_{j}^{\prime
},\theta _{j}^{\prime })$ is deemed possible by $\theta _{i},$ it holds that 
$w_{j}(\theta _{j}^{\prime })\neq u_{j}.$ Since $\theta _{i}$ prior assumes $%
u,$ $\theta _{i}$ deems $(c_{j},\theta _{j})$ infinitely more likely than $%
(c_{j}^{\prime },\theta _{j}^{\prime })$. It follows that $%
t_{i}(E_{i}(\theta _{i}))$ deems $(c_{j},t_{j})$ infinitely more likely than 
$(c_{j}^{\prime },t_{j}^{\prime }).$ Here we have shown that $%
t_{i}(E_{i}(\theta _{i}))$ assumes $j$'s rationality.\smallskip

Suppose that, for some $n\in \mathbb{N},$ we have shown that for each $k\leq
n,$\smallskip \newline
(n1) if $t_{i}\in T_{i}$ expresses $k$-fold assumption of rationality, then
each $\theta _{i}\in \Theta _{i}(t_{i})$ expresses $k$-fold assumption of
prior $u$ and that every good choice is supported;\smallskip \newline
(n2) If $\theta _{i}\in \Theta _{i}$ expresses $k$-fold full belief of
rationality and $k$-fold assumption of prior $u$ and that every good choice
is supported, then $t_{i}(E_{i}(\theta _{i}))$ expresses $k$-fold assumption
of rationality.\smallskip

Now we show that these two statements hold for $n+1.$ First, suppose that $%
t_{i}\in T_{i}$ expresses $(n+1)$-fold assumption of rationality. Let $%
c_{j}\in C_{j}$ be a choice of $j$ optimal for some cautious type whose
assigned utility function is $u_{j}$ that expresses up to $n$-fold
assumption of prior $u$ and that every good choice is supported. Then it is
easy to see that (1) by inductive assumption, in the constructed complete
information model the corresponding type expresses $n$-fold assumption of
rationality, and (2) $c_{j}$ is optimal for that type. Since $t_{i}$
expresses $(n+1)$-fold assumption of rationality, $t_{i}$ deems possible a
cautious type $t_{j}$ that expresses up to $n$-fold assumption of
rationality and for which $c_{j}$ is optimal. By construction, it follows
that $\theta _{i}$ deems possible some $\theta _{j}\in \Theta _{j}(t_{j})$.
By inductive assumption it follows that each $\theta _{j}\in \Theta
_{j}(t_{j})$ expresses $n$-fold assumption of in that every good choice is
supported. Since $\theta _{i}$ expresses common belief in caution and
rationality it follows that $\theta _{i}$ deems $(c_{j},\theta _{j1})$ for $%
\theta _{j1}\in \Theta _{j}(t_{j})$ (that is, $w_{j}(\theta _{j1})=u_{j}$).

Let $(c_{j},\theta _{j})$ with $\theta _{j}$ cautious deemed possible by $%
\theta _{i}$ satisfying that $\theta _{j}$ expresses up to $n$-fold
assumption of prior $u$ and that every good choice is supported and $%
w_{j}(\theta _{j})=u_{j}$ and $(c_{j}^{\prime },\theta _{j}^{\prime })$ a
pair which does not satisfy those conditions. Let $(c_{j},t_{j})$ and $%
(c_{j}^{\prime },t_{j}^{\prime })$ be the pairs occurring in the belief of $%
t_{i}$ corresponding to $(c_{j},\theta _{j})$ and $(c_{j}^{\prime },\theta
_{j}^{\prime }).$ Since $c_{j}$ is rational $\theta _{j}$ with $w_{j}(\theta
_{j})=u_{j},$ it follows that $c_{j}$ is optimal for $t_{j}.$ Also, by
inductive assumption, it follows that $t_{j}$ expresses up to $n$-fold
assumption of rationality. On the other hand, it can be seen that $%
(c_{j}^{\prime },t_{j}^{\prime })$ does not satisfy these conditions. Since $%
t_{i}$ expresses $(n+1)$-fold of assumptions of rationality, $t_{i}$ deems $%
(c_{j},t_{j})$ infinitely more likely than $(c_{j}^{\prime },t_{j}^{\prime
}).$ By construction, $\theta _{i}$ deems $(c_{j},\theta _{j})$ infinitely
more likely than $(c_{j}^{\prime },\theta _{j}^{\prime }).$ Here we have
shown that $\theta _{i}$ expresses $(n+1)$-fold assumption of prior $u$ and
that every good choice is supported.\smallskip

Now suppose that $\theta _{i}\in \Theta _{i}$ expresses $(n+1)$-fold
assumption of prior $u$ and that every good choice is supported. Let $%
c_{j}\in C_{j}$ be a choice of $j$ optimal for some cautious type that
expresses to $n$-fold assumption of rationality. By inductive assumption it
follows that the corresponding type within some incomplete information model
also expresses $n$-fold assumption of prior $u$ and that every good choice
is supported. It can be seen that $c_{j}$ is optimal to the constructed type
having $u_{j}$ as its utility function and the type expresses up to $n$-fold
assumption of prior $u$ and that every good choice is supported. Then $%
\theta _{i}$ deems possible a type $\theta _{j}$ with $w_{j}(\theta
_{j})=u_{j}$ for player $j$ which expresses up to $n$-fold assumption of
prior $u$ and that every good choice is supported for which $c_{j}$ is
optimal. By inductive assumption it follows that $t_{i}(E_{i}(\theta _{i}))$
deems possible $t_{j}(E_{j}(\theta _{j}))$ which expresses $n$-fold
assumption of rationality and for which $c_{j}$ is optimal.

Let $(c_{j},t_{j}$) be a pair with $t_{j}$ cautious deemed possible by $%
t_{i}(E_{i}(\theta _{i}))$ where $t_{j}$ expresses up to $n$-fold assumption
of rationality and $c_{j}$ is optimal for $t_{j}$, and let $(c_{j}^{\prime
},t_{j}^{\prime })$ be a pair that does not satisfy this property. It can be
seen that there is some $\theta _{j}$ corresponding to $t_{j}$ such that $%
\theta _{j}$ is cautious and expresses up to $n$-fold assumption of prior $u$
and that every good choice is supported and $w_{j}(\theta _{j})=u_{j},$while 
$(c_{j}^{\prime },\theta _{j}^{\prime })$ does not satisfy this property for
any $\theta _{j}^{\prime }$ deemed possible by $\theta _{i}$ since $\theta
_{i}$ expresses $n$-fold full belief of rationality. Therefore $%
t_{i}(E_{i}(\theta _{i}))$ deems $(c_{j},t_{j})$ infinitely more likely than 
$(c_{j}^{\prime },t_{j}^{\prime }).$ Here we have shown that $%
t_{i}(E_{i}(\theta _{i}))$ expresses $(n+1)$-fold assumption of rationality. 
$\square $\smallskip \newline
\textbf{Proof of Lemma \ref{sec:cr}.1}. The only-if part holds
automatically. To show the if part, we need first to show that each weak
cautious type can be extended into a cautious one without changing the set
of choices rational for it. It is done by an interpolation method as
follows. Let $t_{i}$ be a type satisfying weak caution with $%
b_{i}(t_{i})=(b_{i1},...,b_{iK})$, $c_{j}\in C_{j},$ and $t_{j}\in
T_{j}(t_{i}).$ Suppose that $(c_{j},t_{j})$ is not deemed possible by $%
t_{i}. $ Since $t_{i}$ is weakly cautious, there is some $t_{j}^{\prime }\in
T_{j}$ such that for some $k\in \{1,...,K\},$ $b_{ik}(c_{j},t_{j}^{\prime
})>0.$ Now we extend $(b_{i1},...,b_{iK})$ into $(b_{i1}^{\prime
},...,b_{i,K+1}^{\prime })$ by letting (1) $b_{it}^{\prime }=b_{it}$ for
each $t\leq k,$ (2) $b_{it}^{\prime }=b_{i,t-1}$ for each $t>k+1,$ and (3) $%
b_{i,k+1}^{\prime }$ is obtained by replacing every occurrence of $%
(c_{j},t_{j}^{\prime })$ by $(c_{j},t_{j})$ in the distribution of
\thinspace $b_{ik}.$ We call $b_{i,k+1}^{\prime }$ a \emph{doppelganger} of $%
b_{ik}.$ It can be seen that for each $c_{i}\in C_{i},$ and a doppelganger $%
b_{i,k+1}^{\prime }$ of $b_{ik},$ $u_{i}(c_{i},b_{i,k+1}^{\prime
})=u_{i}(c_{i},b_{ik}).$ By repeatedly interpolating doppelgangers into $%
b_{i}(t_{i})$ for each missed choice-type pairs, finally we obtain a
lexicographic belief $(b_{i1}^{\prime },...,b_{iK^{\prime }}^{\prime })$
that satisfies caution. We use $\overline{t}_{i}$ to denote the type with
belief $(b_{i1}^{\prime },...,b_{iK^{\prime }}^{\prime })$. $\overline{t}%
_{i} $ is called a \emph{cautious extension} of $t_{i}.$ We have the
following lemma.\smallskip \newline
\textbf{Lemma \ref{sec:pro}.1 (Extended type preserves rational choices). }%
Let $t_{i}$ be a weakly cautious type and $\overline{t}_{i}$ a cautious
extension of $t_{i}.$ Then $c_{i}\in C_{i}$ is rational for $t_{i}$ if and
only if it is rational for $\overline{t}_{i}.$\smallskip \newline
\textbf{Proof. (Only-if)} Suppose that $c_{i}$ is not rational for $%
\overline{t}_{i}.$ Then there is some $c_{i}^{\prime }\in C_{i}$ which is
preferred $c_{i}$ under $b_{i}(\overline{t}_{i})=(b_{i1}^{\prime
},...,b_{iK^{\prime }}^{\prime }),$ that is, there is some $k^{\prime }\in
\{0,...,K^{\prime }\}$ such that $u_{i}(c_{i},b_{i\ell }^{\prime
})=u_{i}(c_{i}^{\prime },b_{i\ell }^{\prime })$ for each $\ell \leq
k^{\prime }$ and $u_{i}(c_{i},b_{i,k^{\prime }+1})<u_{i}(c_{i}^{\prime
},b_{i,k^{\prime }+1}).$ Let $b_{i,k+1}$ be the entry in $b_{i}(t_{i})$ such
that $b_{i,k^{\prime }+1}^{\prime }$ is its doppelganger. It follows that in
the original $b_{i}(t_{i})=(b_{i1},...,b_{iK}),$ $u_{i}(c_{i},b_{i\ell
})=u_{i}(c_{i}^{\prime },b_{i\ell })$ for each $\ell \leq k$ and $%
u_{i}(c_{i},b_{i,k+1})<u_{i}(c_{i}^{\prime },b_{i,k+1}).$ Hence $c_{i}$ is
not rational for $t_{i}.$

\textbf{(If)} Suppose that $c_{i}$ is not rational for $t_{i}.$ Then there
is some $c_{i}^{\prime }\in C_{i}$ which is preferred $c_{i}$ under $%
b_{i}(t_{i})=(b_{i1},...,b_{iK}),$ that is, there is some $k\in \{0,...,K\}$
such that $u_{i}(c_{i},b_{i\ell })=u_{i}(c_{i}^{\prime },b_{i\ell })$ for
each $\ell \leq k$ and $u_{i}(c_{i},b_{i,k+1})<u_{i}(c_{i}^{\prime
},b_{i,k+1}).$ Let $b_{i,k^{\prime }+1}^{\prime }$ be the corresponding
doppelganger in $b_{i}(\overline{t}_{i})$ to $b_{i,k+1}$. It follows that in
the original $u_{i}(c_{i},b_{i\ell }^{\prime })=u_{i}(c_{i}^{\prime
},b_{i\ell }^{\prime })$ for each $\ell \leq k^{\prime }$ and $%
u_{i}(c_{i},b_{i,k^{\prime }+1}^{\prime })<u_{i}(c_{i}^{\prime
},b_{i,k^{\prime }+1}^{\prime }).$ Hence $c_{i}$ is not rational for $%
\overline{t}_{i}.$ $\square $\smallskip \newline
\textbf{Proof of Lemma \ref{sec:cr}.1} \textbf{(Continued)} Since caution
implies weak caution, the only-if part holds automatically. For the if part,
suppose that $c_{i}^{\ast }$ is rational for some $t_{i}^{\ast }\in T_{i}$
which expresses common full belief in weak caution and primary belief in
rationality. Consider an epistemic model $(\overline{T}_{i},\overline{b}%
_{i})_{i\in I}$ such that for each $i\in I,$ $\overline{T}_{i}=\{\overline{t}%
_{i}:t_{i}\in T_{i}\}$ and $\overline{b}_{i}(\overline{t}_{i})$ is a
cautious extension of $b_{i}(t_{i})$ with replacing each occurrence of $%
t_{j} $ by $\overline{t}_{j}$. By Lemma \ref{sec:pro}.1, since $c_{i}^{\ast }
$ is rational for $t_{i}^{\ast },$ it is also rational for $\overline{%
t_{i}^{\ast }}.$ Also, it can be seen by construction that $\overline{%
t_{i}^{\ast }}$ expresses common full belief in caution. Also, since the
interpolation always put doppelgangers after the original one, it does not
change the level-1 belief, and consequently $\overline{t_{i}^{\ast }}$
expresses common full belief in primary belief in rationality. Therefore, $%
c_{i}^{\ast }$ is permissible. $\square$


\begin{thebibliography}{}
%
%
\bibitem{a01} Asheim GB (2001) Proper rationalizability in lexicographic
beliefs. Int J Game Theory 30:453-478.

\bibitem{b03} Battigalli P (2003) Rationalizability in infinite, dynamic
games of incomplete information. Res Econ 57:1-38.

\bibitem{bs03} Battigalli P, Siniscalchi M (2003) Rationalization and
incomplete information. BE J Theor Econ 61:165-184.

\bibitem{bs07} Battigalli P, Siniscalchi M (2007) Interactive
epistemology in games with payoff uncertainty. Res Econ 3:1534-5963.

\bibitem{b84} Bernheim D (1984) Rationalizable strategic behavior. Econometrica 52:1007-1028.

\bibitem{bbd91a} Blume L, Brandenburger A, Dekel E (1991a) Lexicographic
probabilities and choice under uncertainty. Econometrica 59:61-79.

\bibitem{bbd91b} Blume L, Brandenburger A, Dekel E (1991b) Lexicographic
probabilities and equilibrium refinements. Econometrica 59:81-98.

\bibitem{be79} B\"{o}ge W, Eisele T (1979) On solutions of Bayesian
games. Int J Game Theory 8:193-215.

\bibitem{bo94} B\"{o}rgers T (1994) Weak dominance and approximate common
knowledge. J Econ Theory 64:265-276.

\bibitem{bs94} B\"{o}rgers T, Samuelson L (1992) \textquotedblleft
Cautious\textquotedblright\ utility maximization and iterated weak
dominance. Int J Game Theory 21:13-25.

\bibitem{b92} Brandenburger A (1992) Lexicographic probabilities and
iterated admissibility. In Dasgupta J et al. (ed) Economic Analysis of Markets and Games. MIT Press, Cambridge, pp 282-290.

\bibitem{bfk08} Brandenburger A, Friedenberge A, Keisler J (2008)
Admissibility in games. Econometrica 76:307-352.

\bibitem{df90} Dekel E, Fudenberg D (1990) Rational behavior with payoff
uncertainty. J Econ Theory 52:243-267.

\bibitem{ds15} Dekel E, Siniscalchi M (2015) Epistemic game theory. In Young PH, Zamir S (ed) Handbooks of Game Theory with Economic Applications Vol.4. Elsevier, Amsterdam, pp 619-702.

\bibitem{h50} Hamming RW (1950) Error detecting and error correcting
codes. Bell Syst Tech J 29: 147-160.

\bibitem{hs98} Heifetz A, Samet D (1998) Topology-free typology of
beliefs. J Econ Theory 82:324-341.

\bibitem{m78} Myerson, RB (1978) Refinements of Nash equilibrium concept. Int J Game Theory 7:73-80.

\bibitem{p84} Pearce D (1984) Rational strategic behavior and the problem
of perfection. Econometrica 52:1029-1050.

\bibitem{p12} Perea A (2012) Epistemic Game Theory: Reasoning and
Choice. Cambridge University Press, Cambridge.

\bibitem{pk16} Perea A, Kets W (2016) When do types induce the same
belief hierarchy? Games 7. http://dx.doi.org/10.3390/g7040028.

\bibitem{ps17} Perea A, Roy S (2017) A new epistemic characterization of $%
\varepsilon $-proper rationalizability. Games Econ Behav 104:309-328.

\bibitem{s92} Samuelson L (1992) Dominated strategies and common knowledge. 
Games Econ Behav 4:284-313.

\bibitem{sc99} Schuhmacher F (1999) Proper rationalizability and backward
induction. Int J Game Theory 28:599-615.

\bibitem{s75} Selten R (1975) Reexamination of the perfectness concept for
equilibrium points in extensive games. Int J Game Theory 4:25-55.
\end{thebibliography}


\end{document}